\pdfoutput=1
\documentclass{sig-alternate} 

\usepackage{times,graphicx,xspace,comment,subfigure,algorithm,algpseudocode,multirow,url,amsmath}
\usepackage{tabularx}
\usepackage{booktabs}
\usepackage{balance} 

\newcommand{\remove}[1]{}
\newfont{\cour}{cmtt10}
\newcommand{\ol}{\setlength{\itemsep}{0pt.}\begin{enumerate}}
\newcommand{\eol}{\end{enumerate}\setlength{\itemsep}{-\parsep}}
\newcommand{\eg}{\emph{e.g.}\xspace}
\newcommand{\ie}{\emph{i.e.}\xspace}
\newcommand{\etal}{\emph{et al.}\xspace}

\renewcommand{\paragraph}[1]{\vspace{0.1in}\noindent{\bf #1.}}

\newtheorem{definition}{\bf Definition}[section]

\newcounter{proposition}

\makeatletter
{\catcode`\/\active\catcode`\_\active\catcode`\.\active
\gdef\URLslash{\futurelet\next\@@URLslash}%
\gdef\@@URLslash{\ifx\next\URLslash\char`\/\else\slash\fi}%
\gdef\URLdot{\char`\.\penalty\exhyphenpenalty}%
\gdef\URLprepare{\catcode`\/\active\catcode`\_\active\catcode`\.\active
        \let/\URLslash\let.\URLdot\def~{\char`\~}\def_{\char`\_}}}%
\def\URL{\bgroup\URLprepare\realURL}%
\def\realURL#1{\tt #1\egroup}%

\newcommand{\secref}[1]{\mbox{Section~\ref{#1}}}

\begin{document}

\conferenceinfo{BigMine}{'12, August 12, 2012 Beijing, China}

\title{Space-Efficient Sampling from Social Activity Streams}

\numberofauthors{3} 

\author{
Nesreen K. Ahmed, Jennifer Neville, and Ramana Kompella\\
       \affaddr{Computer Science Department, Purdue University}\\
       \email{\{nkahmed,neville,kompella\}@cs.purdue.edu}
}

\maketitle
\begin{abstract}
In order to efficiently study the characteristics of network
domains and support development of network systems (\eg algorithms, protocols
that operate on networks), it is often necessary to {\em sample} a
representative subgraph from a large complex network.  Although recent subgraph sampling methods have been shown to work well, they focus on sampling from memory-resident graphs and assume that the sampling algorithm can access the entire graph in order to decide which nodes/edges to select. Many large-scale network datasets, however, are too large and/or dynamic to be processed using main memory (e.g., email, tweets, wall posts). 
In this work, we formulate the problem of sampling from large graph streams. 
We propose a streaming graph sampling algorithm that dynamically maintains a representative sample in a reservoir based setting. 
We evaluate the efficacy of our proposed methods empirically using several real-world data sets. Across all
datasets, we found that our method produce samples that preserve better the original graph
distributions.  
\end{abstract}

\section{Introduction}
\label{sec:intro}

Many real-world complex systems can be represented as graphs
and networks---from information networks, to communication
networks, to biological networks. Naturally, there has been
a lot of interest in studying characteristics of these
networks, modeling their structure, as well as developing
algorithms and systems that operate on the networks. While
the recent surge in activity in online social networks (\eg,
Facebook, Twitter) has prompted a similar need for
characterization and modeling efforts, it is often much
harder than in traditional networks due to their size.
Specifically, these networks tend to be too large to
efficiently acquire, store and/or analyze (\eg, one billion
chat messages per day in Facebook~\cite{facebookstats}).
It is therefore often necessary to {\em sample} smaller
subgraphs from the larger network structure, that 
can then be used to investigate the characteristics and
properties of the larger network.  It can also be used to
drive realistic simulations and experimentation before
deploying new protocols and systems in the field---for
example, new Internet protocols, social/viral marketing
schemes, and/or fraud detection algorithms. 

In this work, we consider the following graph sampling
problem: Assume an input graph $G\!=\!(V,E)$ of size
$N=|V|$, from which a sampling algorithm selects a subgraph
$G_s\!=\! (V_s, E_s)$ with a subset of the nodes ($V_s
\!\subset\! V$) and/or edges ($E_s \!\subset\! E$), such
that $|V_s|=\phi N$.  We refer to $\phi$ as the sampling
fraction.  The goal is to sample a {\em representative}
subgraph $G_s$ that matches many of the properties of $G$,
so that $G_s$ can be used to {\em simultaneously} preserve
several characteristics of the network structure in the
original graph $G$ (\eg, degree, path length, clustering).
Specifically, we aim to select a $G_s$ that minimizes the
distributional distance over several graph measures (\eg,
degree distribution) simultaneously. Let $f(.)$ be a
property of a graph, then the goal is to select a sample
that minimizes the distance between the property in $G$ and
the property in $G_s$: $dist[ f(G),f(G_S)]$. In this work,
we consider degree, hop plot, and clustering distributions
for $f(.)$ and evaluate using two distributional distance
metrics---Kolmogorov-Smirnov distance and skew
divergence~\cite{lee:01}.

While many graph sampling methods have been proposed before
(\eg, \cite{hubler08icdm,leskovec2006slg}), 
they typically require access to the whole graph in its entirety at any step, in order to
decide which nodes/edges to select.  While the graph data
can be stored on disks, processing full large graphs is
usually done using physical memory (RAM) which is a
limited/expensive resource.  Therefore, the ideal approach
to process the graph data is to use a {\em streaming model},
where the graph data is presented as a stream of edges, and
any computation on the stream relies on using a small amount
of memory, and in a single pass.  Many large-scale network
datasets readily admit such a streaming model.  For example,
online social network applications (\eg Facebook, Twitter)
have data that consist of micro-communications among users
(\eg wall posts, tweets, emails); any activity between two
users can result in an edge getting added to the activity
graph. 

In this work, we consider the problem of sampling from such
large social activity streams. We refer to social activity streams as graph streams since social activities can be represented as a graph.
Specifically, {\em our goal} is to devise a streaming
algorithm for sampling subgraphs from large graph streams,
that can decide whether to include an edge in the sampled
graph, as the edge is streamed in.

While there is a great deal of research on data streams and
data stream management, to our knowledge, our work is the first
work to focus on streaming algorithms for sampling subgraphs
from large graph streams in a single pass.  Satisfying the dual objective of
finding a sampling algorithm that can sample representative
subgraphs, while being amenable to a streaming
implementation is quite challenging. Most existing sampling algorithms fail to process graph streams (\ie requiring multiple passes over the edges). As an example, breadth-first search needs to access the full neighborhood of a node to perform one step of its process.

In this paper, we propose a novel sampling algorithm that is amenable to streaming implementations. Specifically, we propose  {\em partially-induced edge sampling (PIES)} that randomly samples edges, induces the sampled nodes, and maintains a dynamic/changing sample in a reservoir-based setting using a single pass over the edges.    

Our proposed approach is simple, efficient, and can be used to sample large graphs that are too large to fit in memory. Moreover, it can also be used to graphs that readily admit the streaming model (\eg email logs, tweets between users in Twitter).

We evaluate PIES over a number of real world (\eg,
Facebook, Twitter, HepPH, Flickr) datasets collected by other
researchers (\cite{leskovecRepository,viswanath-2009-activity}), and an
email network constructed from two weeks of Purdue
University email traffic. We compare PIES to existing/proposed baseline
stream sampling techniques such as edge sampling, node
sampling and a simple breadth-first search (BFS) based
algorithm. 

Across all datasets, we observed that PIES
produces samples that better match the distributions of
degree, path length and clustering compared to other
existing algorithms.

The rest of the paper is organized as follows. We first present a background and related work on
sampling methods in section~\ref{sec:background} . Next, we outline our proposed
sampling algorithms with streaming implementations in section~\ref{sec:approach}.
Finally, we compare PIES with other baseline graphs sampling algorithms in section~\ref{sec:experimental-analysis}.

\section{Graph Sampling Algorithms}
\label{sec:background}

In this section, we discuss standard graph sampling
algorithms in literature, which can be broadly classified as
node-based, edge-based, and topology-based methods.  Most
graph sampling algorithms have two basic components: (1)
node selection, and (2) subgraph formation. The node
selection step identifies a sample set of nodes ($V_s$),
while the subgraph formation step selects the set of edges
($E_s$) to be included in the sampled subgraph. We distinguish
between two different approaches to subgraph formation---total
and partial graph induction---which differ by whether {\em
all} or {\em some} of the edges incident on the sampled
nodes are included in the sampled graph.  The resulting
sampled graphs are referred to as the {\em induced subgraph}
and {\em partially induced subgraph} respectively.  

\paragraph{Node sampling (NS)} In classic node sampling,
nodes are chosen independently and uniformly at random from
the original graph for inclusion in the sampled graph. For a
target fraction $\phi$ of nodes required, each node is
simply sampled with a probability of $\phi$.  Once the nodes
are selected, the sampled graph consists of the {\em induced
subgraph} over the selected nodes, \ie, all edges among the
sampled nodes are added to form the sampled graph.
While node sampling is intuitive and relatively
straightforward, the work in \cite{stumpf2005ssf} shows that
it does not accurately capture properties for graphs with
power-law degree distributions.  Similarly, \cite{lee:06}
shows that although node sampling appears to capture nodes
of different degrees well, due to its inclusion of all edges
for a chosen node set, the original level of connectivity is
not likely to be preserved.  

\paragraph{Edge sampling (ES)} Edge sampling focuses on the
selection of edges rather than nodes to populate the sample.
Thus, the node selection step in edge sampling algorithm
proceeds by just sampling edges, and including both nodes
when a particular edge is sampled.  The partially induced
graph is created just out of the sampled edges, which means
no extra edges are added in addition to those chosen during
the random edge selection process. 
Unfortunately, ES fails to preserve many desired graph properties due to the
independent sampling of edges.
It is however more likely to capture path lengths, due to its bias towards high
degree nodes and the inclusion of both end points of selected edges. 

\paragraph{Topology-based sampling} Due to the known
limitations of NS (\cite{stumpf2005ssf,lee:06}) and ES (bias
toward high degree nodes), researchers have also considered
many other topology-based sampling methods. One example is
snowball sampling, which selects nodes using breadth-first
search from a randomly selected seed node.  Snowball
sampling accurately maintains the network connectivity
within the snowball, however it suffers from a
\emph{boundary bias} in that many peripheral nodes (\ie,
those sampled on the last round) will be missing a large
number of neighbors~\cite{lee:06}.
In~\cite{leskovec2006slg}, Leskovec \etal propose a
Forest Fire Sampling (FFS) method. 
It starts by picking a node uniformly at random, then`burns' a random fraction of its outgoing
links. The process is recursively repeated until no new node is selected or we obtain
the sample size. In general, such topology-based sampling approaches perform better than NS and ES. 

\smallskip None of the algorithms discussed above have been
explicitly designed to work in a streaming fashion, as the
emphasis has been largely on sampling representative
subgraphs that matched the properties of the original graph
well. In the next section, we discuss our model of graph
streams, and show how these standard sampling algorithms
could be adapted to work in such a streaming setting. We
also propose our new algorithm that outperforms simple
streaming variants of these algorithms in the next section. 

\section{Stream Sampling}
\label{sec:approach}

We consider an undirected graph $G(V,E)$ with a vertex set $V=\{v_1,v_2,...,v_N\}$ and edge set $E=\{e_1,e_2,...,e_M\}$ where $N$ is the number of vertices and $M$ is the number of edges in $G$. We assume $G$ arrives as a graph stream.
\noindent 
\begin{definition} We define a \emph{graph stream} as a sequence of edges 
$e_{\pi(1)},e_{\pi(2)},...,e_{\pi(M)}$, where $\pi$ is any random permutation on $[M]=\{1,2,...,M\}$, 
$\pi : [M] \rightarrow [M]$.   
\label{gstream_defn}
\end{definition}

In traditional computational models of graphs, it is difficult
to perform random access of the entire graph $G$ at any
step, since it is unlikely for large graphs to easily fit in
the main memory. 
A streaming model, in which the graph can only be accessed as
a stream of edges, arriving one edge at a time, is therefore more preferable \cite{Zhang10}.

In a streaming model, 
as each edge $e \in E$ arrives, the sampling algorithm $\sigma$ needs to decide whether
to include the edge or not as the edge is {\em streamed} in.  The sampling
algorithm $\sigma$ may also maintain state $\Psi$, and consult the state to
determine whether to sample a subsequent edge or not, but the total storage
associated with $\Psi$ should be of the order the size of the output sampled
graph $G_s$, \ie, $|\Psi|$ = $O(|G_s|)$.  Note that this requirement is
potentially larger than the $o(N,t)$ (preferably, $polylog(N,t)$) that streaming
algorithms typically require~\cite{muthu}. But, since any algorithm cannot
require less space than the output, we relax this requirement in our
definition as follows.  

\begin{definition} We define a \emph{streaming graph
sampling algorithm} as any sampling algorithm $\sigma$ that
produces a sampled graph $G_s$ such that ${\vert
V_s\vert}/{\vert V \vert}=\phi$, which (1) samples edges of
the original graph $G(V,E)$ in a sequential order (\ie, not
random access) in {\em one pass}; and,  (2) maintains state
$\Psi$ that is of the order of the size of the sampled graph
$G_s$, \ie, $|\Psi|=O(|G_s|)$.  \label{stream_defn}
\end{definition}
 
Now, using the above definition of a streaming graph
sampling algorithm, we now present streaming variants of
different algorithms discussed in \secref{sec:background}.

\subsection{Streaming Node Sampling} One key problem with
traditional node sampling we discussed in
\secref{sec:background} is that nodes are selected at
random. In our stream setting, new nodes arrive into the
system only when an edge that contains the new node is added
into the system; it is therefore hard to identify which $n$
nodes to select {\em a priori}. To address this, we
essentially use the idea of reservoir
sampling~\cite{Vitter:85} and
propose the following streaming node sampling variant
(outlined in Algorithm~\ref{algo:NS}). 

The main idea is to select nodes uniformly at random with
the help of a uniform random hash function. Specifically, we keep track of
nodes with $n$ smallest hash values in the graph; nodes are
only added if their hash values represent the top-$n$
minimum hashes among all nodes seen thus far in the stream.
Any edge that has both vertices already in the reservoir is
automatically added to the original graph. 
Since the reservoir is finite, it can happen that a node
that arrives much later may have a smaller hash value, in
which case it replaces an existing node. All edges incident
on that node are then removed from the sampled graph, as
there is no chance for those edges to ever get sampled
again. Thus, once the reservoir is filled up to $n$ nodes,
it will remain at $n$ nodes, but at the same time, it will
guarantee sampling from all portions of the stream (not just
the front) since the selection is based on the hash value.

\begin{algorithm}[h]
\floatname{algorithm}{Algorithm}
\begin{algorithmic}[1]
\State $\triangleright$ $V_s = \emptyset, E_s = \emptyset$
\State $\triangleright$ $h$ is fixed uniform random hash function
\State $\triangleright$ $t=1$
\For{$e_t$ in the graph stream $S$}
       \State $\triangleright$ $(u, v) = e_t$
       \If {$u \notin V_s \&$ $h(u)$ is top-n min hash}
		\State $V_s = V_s \cup u$
		\State Remove all edges incident on replaced node
       \EndIf
       \If {$v \notin V_s \&$ $h(v)$ is top-n min hash}
		\State $V_s = V_s \cup v$
		\State Remove all edges incident on replaced node 
       \EndIf
       \If {$u,v \in V_s$}
  		\State $E_s = E_s \cup e_t$
	\EndIf 
        \State $\triangleright$ $t=t+1$ 
\EndFor
\State Output $G_s = (V_s, E_s)$
\end{algorithmic}
\caption{Streaming NS(Sample Size $n$, Stream $S$)}
\label{algo:NS}
\end{algorithm}

\subsection{Streaming Edge Sampling} Streaming edge sampling is a simple variant of the traditional edge sampling. Here, instead of
hashing individual nodes, we focus on using hash-based
selection of edges (as shown in Algorithm~\ref{algo:ES}).
More precisely, if we are interested in obtaining $m$ edges
at random from the stream, we can simply keep a reservoir of
$m$ edges with the minimum hash value.  Thus, if a new edge
streams into the system, we check if its hash value is
within top-$m$ minimum hash values. If it is not, then we do
not select that edge, otherwise we add it to the reservoir
while replacing the edge with the previous highest top-$m$
minimum hash value. A similar approach has been proposed by Aggarwal in ~\cite{aggarwal2011outlier}. However, in his work the goal was to get efficient structural compression of the underlying graph stream rather than getting a representative subgraph that can be used instead of the full graph.
One problem with this approach is that our goal is often in
terms of sampling a certain number of nodes $n$. Since we use a
reservoir of edges, finding the right $m$ that provides $n$
nodes is really hard. It also keeps varying depending on
which edges the algorithm ends up selecting. Note that
sampling fraction could also be specified in terms of
fraction of edges; the choice of defining it in terms of
nodes is somewhat arbitrary in that sense. For our
comparison purposes, we ensured that we choose a large
enough $m$ such that the number of nodes was much higher
than $n$, but later iteratively pruned out sampled edges
with the maximum hash values until the target number of
nodes $n$ was reached. While this is not strictly an elegant
streaming algorithm, as we shall show in our evaluation,
even this extra complexity does not result in producing
good graph samples anyway. We include it mainly for
comparison purposes. 

\begin{algorithm}[h]
\floatname{algorithm}{Algorithm}
\begin{algorithmic}[1]
\State $\triangleright$ $V_s = \emptyset, E_s = \emptyset$
\State $\triangleright$ $h$ is fixed uniform random hash function
\State $\triangleright$ $t=1$
\For{$e_t$ in the graph stream $S$}
       \State $\triangleright$ $(u, v) = e_t$
	\If {$h(e_t)$ is in top-$m$ min hash}
            \State $E_s= E_s \cup e_t$ 
	     \State $V_s = V_s \cup \{u,v\}$
        \EndIf
        \State Iteratively remove edges in $E_s$ such that
$n$ nodes.  
        \State $\triangleright$ $t=t+1$ 
\EndFor
\State Output $G_s = (V_s, E_s)$
\end{algorithmic}
\caption{Streaming ES(Sample Size $n$, Stream $S$)}
\label{algo:ES}
\end{algorithm}

\subsection{Streaming Topology-Based Sampling} We also
consider a streaming variant of a topology-based sampling
algorithm. Specifically, we consider a simple BFS-based
algorithm (shown in Algorithm~\ref{algo:BFS}) that works as follows. This algorithm essentially
implements a simple breadth-first search on a sliding window
of $w$ edges in the stream. In many respects, this algorithm
is similar to the forest-fire sampling (FFS) algorithm. Just
as in FFS, it essentially starts at a random node in the
graph and selects an edge to burn (as in FFS parlance) among
all edges incident on that node within the sliding window. 
For every edge burned, let $v$ be the other end of the burned edge. 
We enqueue $v$ onto a queue $Q$ in order to get a chance to burn its incident edges within the window.
For every new streaming edge, the sliding window moves one step,
which means the oldest edge in the window is dropped and a
new edge is added. (If that oldest edge was sampled, it will
still be part of the sampled graph.) 
If as a result of the sliding window moving one step, the
node has no more edges left to burn, then the burning
process will dequeue a new node from $Q$. If the queue is empty, the process jumps to a random node within the sliding window
(just as in FFS). This way, it does BFS as much as possible
within a sliding window, with random jumps if there is no
more edges left to explore.  Note that there may be other
streaming variants of the sampling algorithm possible; since
there are no streaming algorithms in the literature, we
chose this as a reasonable approximation for comparison. 
This algorithm has a similar problem as the edge sampling
variant that it is difficult to control the exact number of
sampled nodes, and hence some additional pruning needs to be
done at the end (as shown in Algorithm~\ref{algo:BFS}).

\begin{algorithm}[h]
\floatname{algorithm}{Algorithm}
\begin{algorithmic}[1]
\State $\triangleright$ $V_s = \emptyset, E_s = \emptyset$
\State $\triangleright$ $W = \emptyset$
\State $\triangleright$ Add the first $wsize$ edges to $W$
\State $\triangleright$ $t=wsize$
\State $\triangleright$ Create a queue $Q$
\State $\triangleright$ $u=$random vertex from $W$
\For{$e_t$ in the graph stream $S$}
	\State \textbf{if} $u \notin V_s$ \textbf{then} add $u$ to $V_s$
	\If {$W.incident\_edges(u) \neq \emptyset$}
		\State Sample $e$ from $W.incident\_edges(u)$
		\State Add $e=(u, v)$ to $E_s$
		\State Remove $e$ from $W$
		\State Add  $v$ to $V_s$
		\State enqueue $v$ onto $Q$
	\Else 
		\State \textbf{if} $Q = \emptyset$ \textbf{then} $u=$random vertex from $W$
		\State \textbf{Else} $u = Q.dequeue()$
	\EndIf 
       \State Move the window $W$
       \If {$|V_s|$ \textgreater $ n$}
  		\State Retain $[e] \subset E_s$ such that $[e]$ has $n$ nodes
		\State Output $G_s = (V_s, E_s)$
	\EndIf 	
       \State $\triangleright$ $t=t+1$ 
\EndFor
\State Output $G_s = (V_s, E_s)$
\end{algorithmic}
\caption{Streaming BFS(Sample Size $n$, Stream $S$,Window Size=$wsize$)}
\label{algo:BFS}
\end{algorithm}

\subsection{Partially-Induced Edge Sampling (PIES)}

We finally present our main algorithm called PIES that
outperforms the above classes of streaming algorithms.  In
our approach, we mainly exploit the observation that edge
sampling is inherently biased towards the selection of nodes
with higher degrees, resulting in an {\em upward bias} in
the degree distributions of  sampled nodes compared to nodes
in the original graph \cite{ribeiro10imc}.  However, in all
sampled subgraphs, degrees are naturally underestimated
since only a fraction of neighbors may be selected. This
results in a {\em downward bias}, regardless of the actual
sampling algorithm used. We also observe that selecting
nodes with high degrees results in samples with higher
average clustering coefficient and shorter path lengths. It
is likely that two interconnected sampled nodes will have
the same neighbor if this neighbor is sampled and has an
extremely large degree. Additionally, high-degree nodes are
usually highly popular in the graph, they serve as good
navigators through the graph and the shortest path is
usually through those extremely popular ones. Therefore,
sampling the high degree nodes can result in connected
sampled subgraphs that accurately preserve the properties of
the full graph.

However, by sampling edges independently, it is
unlikely that the structure of the graph {\em surrounding}
the high degree nodes will be preserved. Thus, we also sample
all the edges between any sampled nodes in the
graph (graph induction). This helps to recover much of the connectivity around the high
degree nodes---offsetting the downward degree bias as well
as increasing local clustering in the sampled graph. Graph
induction increases the probability that triangles will be
formed among the set of sampled nodes, resulting in a higher
clustering coefficient and shorter path lengths.      The
above observations, while simple, makes the sampled graphs
approximate the characteristics of the original graph much
more accurately, even better than topology-based sampling
algorithms. 

\begin{algorithm}[h]
\floatname{algorithm}{Algorithm}
\begin{algorithmic}[1]
\State $\triangleright$ $V_s = \emptyset, E_s = \emptyset$
\State $\triangleright$ $t=1$
\While{graph is streaming}
        \State $\triangleright$ $(u, v) = e_t$,
        \If {$|V_s|$ \textless $ n$}
        	\State \textbf{if} $u \notin V_s$ \textbf{then} $V_s = V_s \cup \{u\}$ 
	\State \textbf{if} $v \notin V_s$ \textbf{then} $V_s = V_s \cup \{v\}$
		\State $E_s = E_s \cup \{e_t\}$		
	\Else
		\State $\triangleright$ $p_{e}=\frac{|E_s|}{t}$
		\State draw $r$ from continuous Uniform(0,1)
		\If {$r \leq p_{e}$}
			\State draw $i$ and $j$ from discrete Uniform[1,$|V_s|$]
        		\State \textbf{if} $u \notin V_s$ \textbf{then} $V_s = V_s \cup \{u\}$ , drop node $V_s[i]$ with all its incident edges
		\State \textbf{if} $v \notin V_s$ \textbf{then} $V_s = V_s \cup \{v\}$ , drop node $V_s[j]$ with all its incident edges
		\EndIf 	
		\State \textbf{if} $u \in V_s$ AND $v \in V_s$  \textbf{then} $E_s = E_s \cup \{e_t\}$
	        \EndIf  
         \State $\triangleright$ $t=t+1$ 
\EndWhile    	
\State Output $G_s = (V_s, E_s)$
\end{algorithmic}
\caption{PIES(Sample Size $n$, Stream $S$)}
\label{algo:pies}
\end{algorithm}

Unfortunately, full graph induction in a streaming fashion
is hard (\ie since it requires
at least two passes, when done in the obvious
straightforward way).  Thus, instead of total induction of
the edges between the sampled nodes, we can utilize {\em
partial} induction and combine the edge-based node sampling
with the graph induction (as shown in
Algorithm~\ref{algo:pies}) into a single step. The partial
induction step induces the sample in the forward direction,
\ie, adding any edge among a pair of sampled nodes if it
occurs after both the two nodes were added to the sample.

PIES aims to maintain a dynamic sample as the graph is
streaming utilizing the same reservoir sampling idea we have
used before.  Specifically, we add the first $n$ records of
the stream to a {\em reservoir} and then the rest of the
stream is processed randomly by replacing existing records
in the reservoir.  PIES will then simply run over the edges
in a single pass, adding deterministically the first $n$
nodes of the stream to the sampled graph.  Once it achieves
the target sample size, then for any streaming edge, it adds
the incident nodes to the sample (probabilistically) by
replacing other sampled nodes from the node sample set
(uniformly at random).  At each step, it will also add the
edge if its two incident nodes are already in the sampled
node set (to produce a partial induction effect). 

\begin{table*} 
\small
\centering 
\begin{tabularx}{\textwidth}{lXXXXXX}
\toprule 
\textbf{{Dataset}} & \textbf{{Nodes}} & \textbf{{Edges}} & \textbf{{No. CC}} & \textbf{{Avg. path}} & \textbf{{Density}} & \textbf{{Clustering}}\\
\midrule
\textsc{HepPH} & 34,546 & 420,877 & 61 & 4.33 & $7\times10^{-4}$ & 0.146\\
\textsc{Twitter} & 8,581 & 27,889 & 162 & 4.17 & $7\times10^{-4}$ & 0.061\\
\textsc{Facebook (NO)} & 46,952 & 183,412 & 842 & 5.6  & $2\times10^{-4}$ & 0.085\\
\textsc{CondMAT} & 23,133 & 93,439 & 567 & 5.35 & $4\times10^{-4}$ & 0.264\\
\textsc{Email-PU Univ} & 214,893 & 1,270,285 & 24 & 3.91 & $5.5\times10^{-5}$ & 0.0018\\
\textsc{Flickr} & 820,878 & 6,625,280 & 1 & 5.01 & $1.9\times10^{-5}$& 0.116\\
\bottomrule
\end{tabularx}
\caption{Characteristics of Network Datasets}
\label{tab:point stats}
\end{table*} 

\section{Experimental Evaluation}
\label{sec:experimental-analysis}
\vspace{1.mm}
In this section, we evaluate the efficacy of the proposed stream sampling algorithms, PIES, NS, ES, and BFS, on several real datasets ranging from about 10,000 - 800,000 nodes, with 30,000 - 6.6 million edges.  
In our experiments, we consider five real networks: a
citation network, a collaboration network, an email communication network, and
two online social networks. 
Table 1 summarizes the characteristics of the (simplified) real networks.

The two data sets called HepPH, and CondMAT, correspond to a citation
graph, and collaboration graph respectively, provided by Leskovec
{\emph et al.} \cite{leskovecRepository}. The Facebook data corresponds to Wall
communications among users that belong to a New Orleans city ~\cite{viswanath-2009-activity}. The Twitter dataset contains tweets of
users in discussion surrounding the United Nations climate change conference in
Dec. 2009. Also, the University email data corresponds to two weeks of email communication we collected
from the email logs on Purdue university mailserver(s). We also verify our proposed approach on a large scale graph of 800,000 nodes and 6.6 million edges collected from Flickr network~\cite{PURR1002}.

\vspace{-1.mm}
\subsection{Evaluation Measures}

We compare four stream sampling methods from different sampling classes. We propose a one-pass implementation of node sampling (NS) and edge sampling (ES) to represent node-based sampling and edge-based sampling classes respectively. We also propose a one-pass breadth first sampling (BFS) to represent the topology-based sampling class. We implement BFS on a sliding window of 100 edges of the stream. Our evaluation is primarily along four main properties---degree, path length,
clustering coefficient, and size of weakly connected components. We conjecture these four properties capture both local and global characteristics of the graph. We measure the performance of a sampling algorithm by how well the sampled subgraphs preserve
the probability density function (PDF) and complementary cumulative distribution function (CCDF) of each of these four properties. Unlike
other measure based on aggregate statistics (\eg, average degree, density, reciprocity), these four measures represent the distribution of properties across the nodes and edges in the sample, which facilitates detailed comparison and evaluation of sample representativeness.

In addition to visually comparing the similarity of the distributions on the sampled subgraphs to those of the original graphs, we also
compute two statistics to compare the distributions quantitatively across different sampling fractions. First, we use the \break Kolmogorov-Smirnov (KS) statistic to assess the distance between two cumulative distribution functions (CDF). The KS-statistic is a widely used measure of the
agreement between two distributions; the authors of \cite{leskovec2006slg} also
have used the KS distance to illustrate the accuracy of FFS samples in the past.
It is computed as the maximum absolute distance between the two distributions, where $x$ represents the range of the random
variable and $F_1$ and $F_2$ represent two CDFs. In this work, $F_1$ represents the true distribution of the full graph and $F_2$ represents the approximation of $F_1$ calculated from the sampled subgraph.
\begin{equation}
KS(F_1,F_2) = max_x{|F_1(x) - F_2(x)|}
\end{equation}
We also used another statistical measure for evaluation called skew divergence, in order to measure the Kullback-Leibler (KL) divergence between two distributions that do not have the same continuous support over the range of values~\cite{lee:01}. The results of skew divergence are similar to the KS statistic results, therefore we omitted them to save space.

\vspace{-1.mm}
\subsection{Results}

In our experiments, we focus on obtaining a sample between 5--40\% ($\phi=0.05$
to $0.40$) of the full graph. For each sample fraction, we experiment with ten different
runs, and in each run, we generate a sample from a new random seed.  It is unlikely to assume a certain order of edges in the stream because usually social communication among users can happen in any arbitrary order. To simulate this aspect we randomly permute the edges in the graph in each run. 

\begin{figure*}[t!]
\centering
\subfigure[Degree]{\label{fig:avg ks deg}\includegraphics[width=0.265\textwidth]{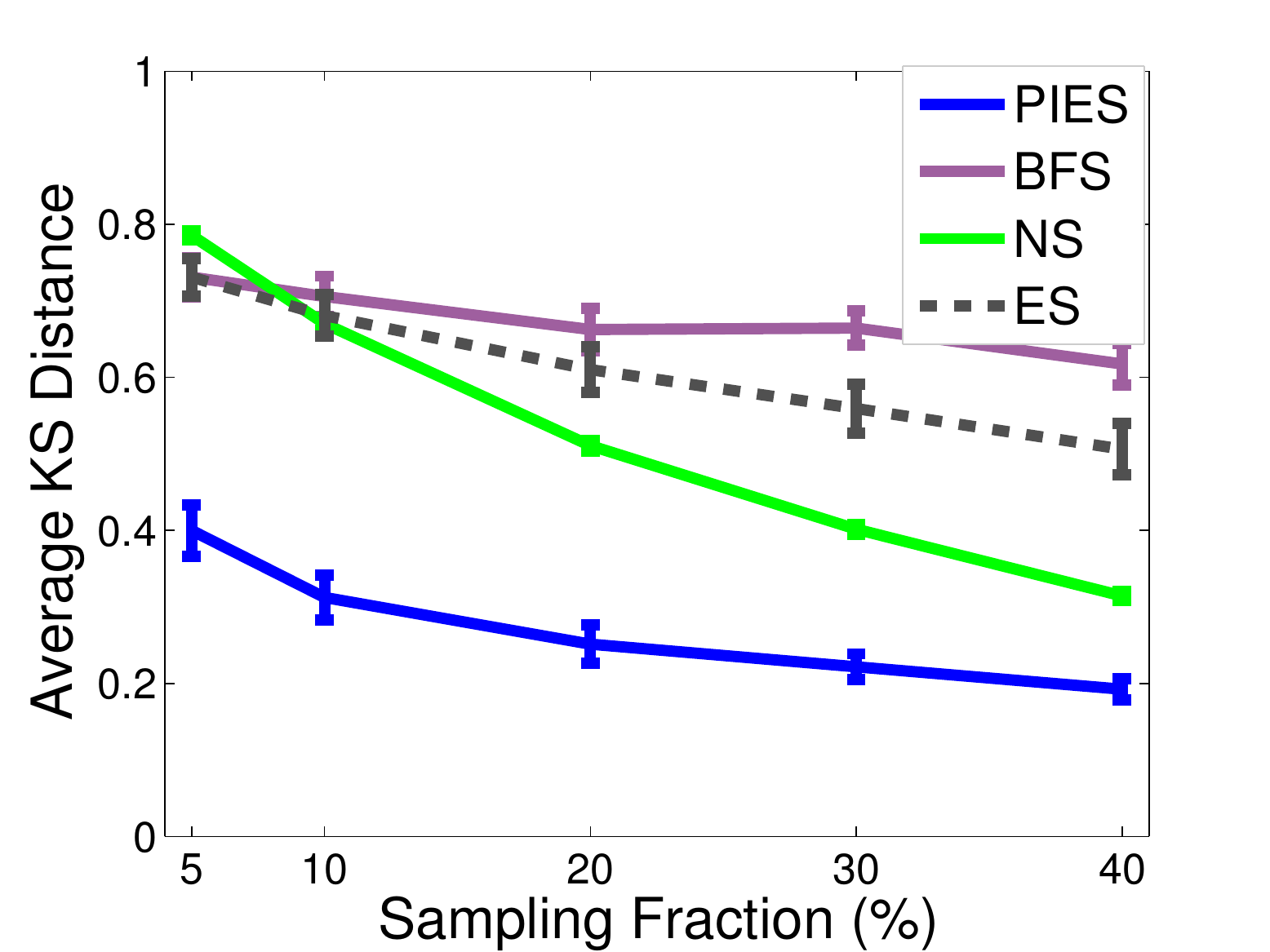}}
\hspace{-5.mm}
\subfigure[Path length]{\label{fig:avg ks plen}\includegraphics[width=0.265\textwidth]{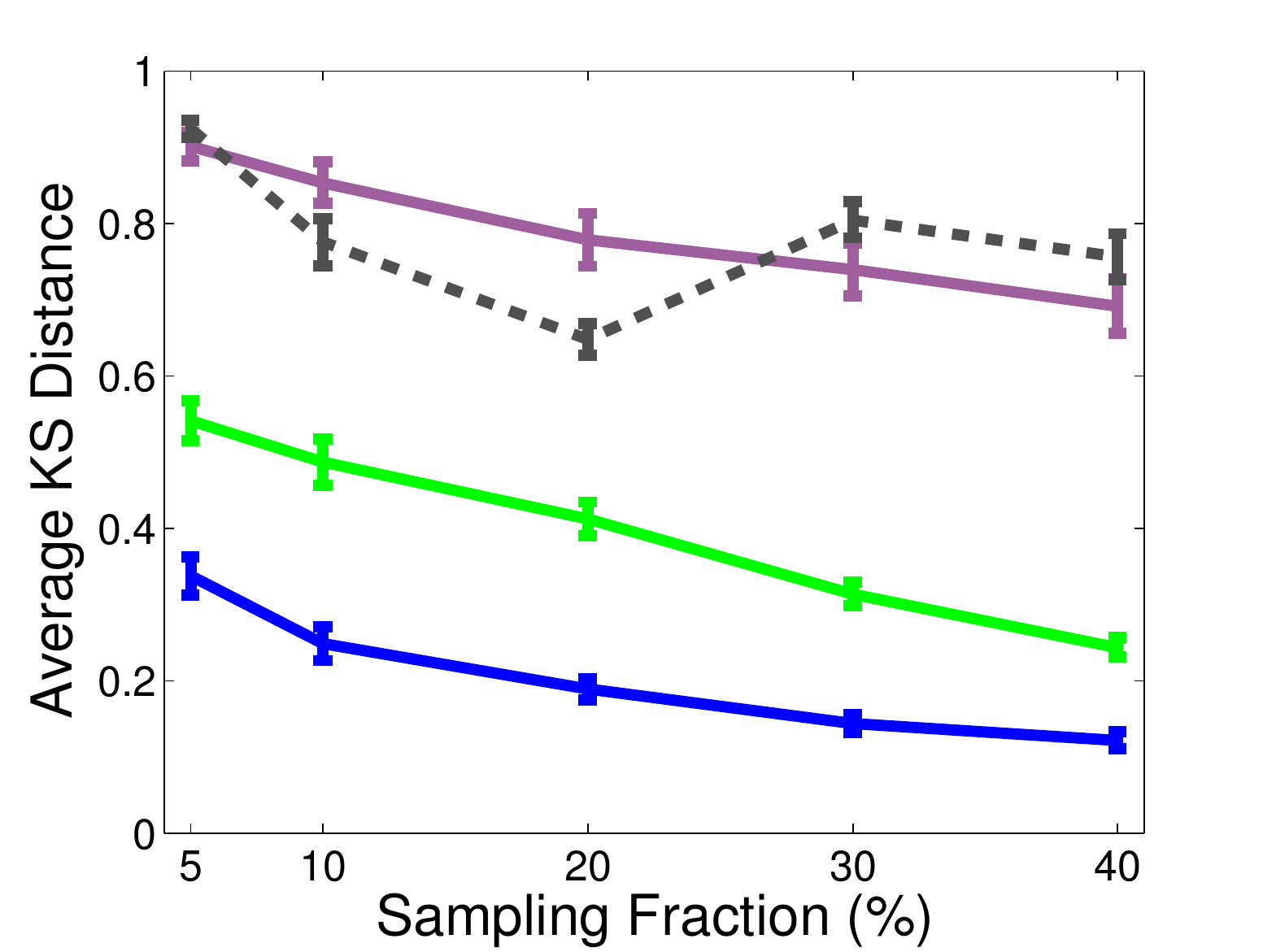}}
\hspace{-5.mm}
\subfigure[Clustering Coefficient]{\label{fig:avg ks cc}\includegraphics[width=0.265\textwidth]{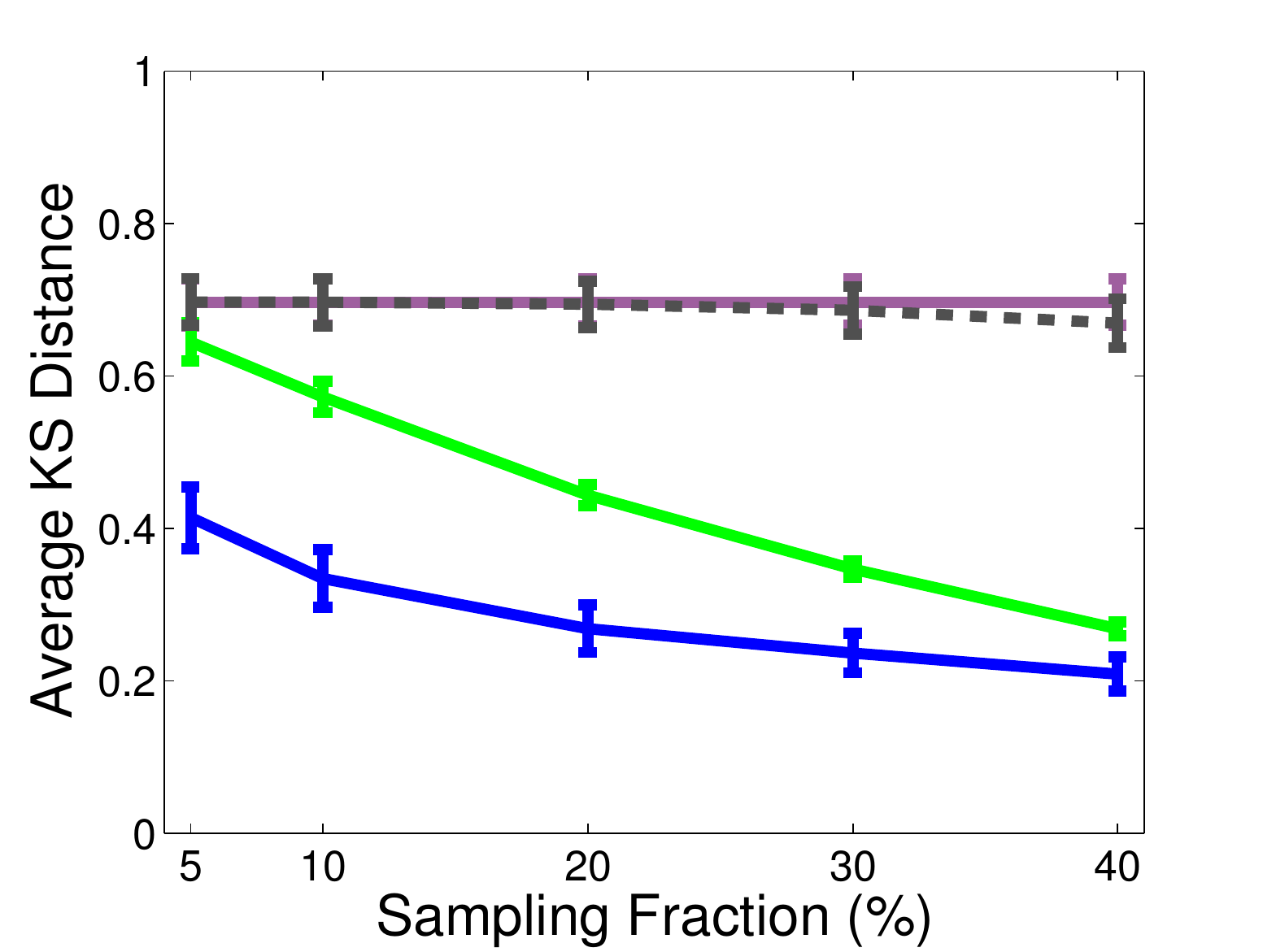}}
\hspace{-5.mm}
\subfigure[Size of Weakly Conn. Comp.]{\label{fig:avg ks comp}\includegraphics[width=0.265\textwidth]{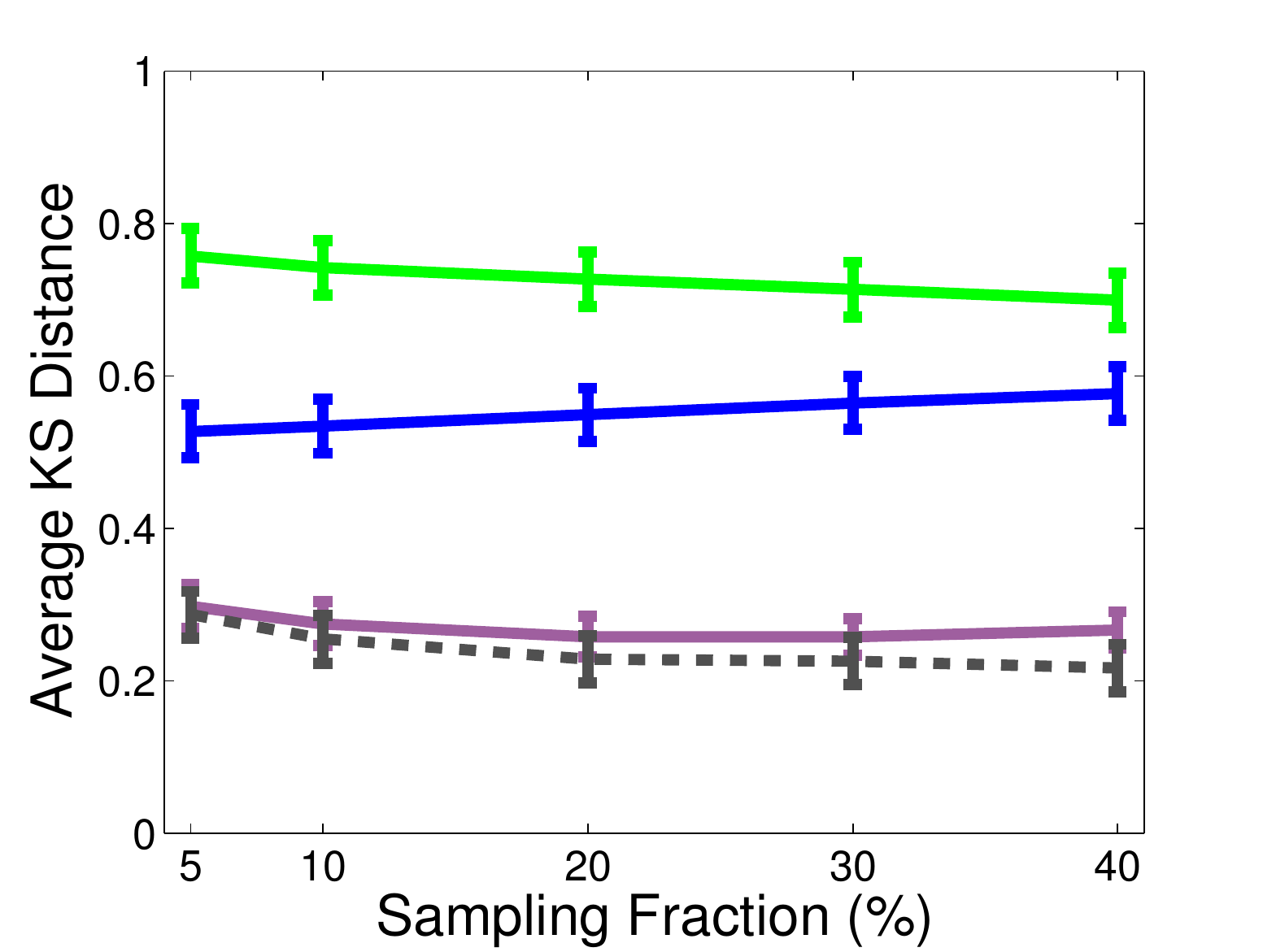}}\\
\vspace{-2mm}
\caption{Average KS Distance across 5 datasets.}
\label{fig:compare_sd}
\end{figure*}

\begin{figure*}[t!]
\centering
\vspace{-2.mm}
\subfigure[Degree]{\label{fig:fbor deg dist}\includegraphics[width=0.265\textwidth]{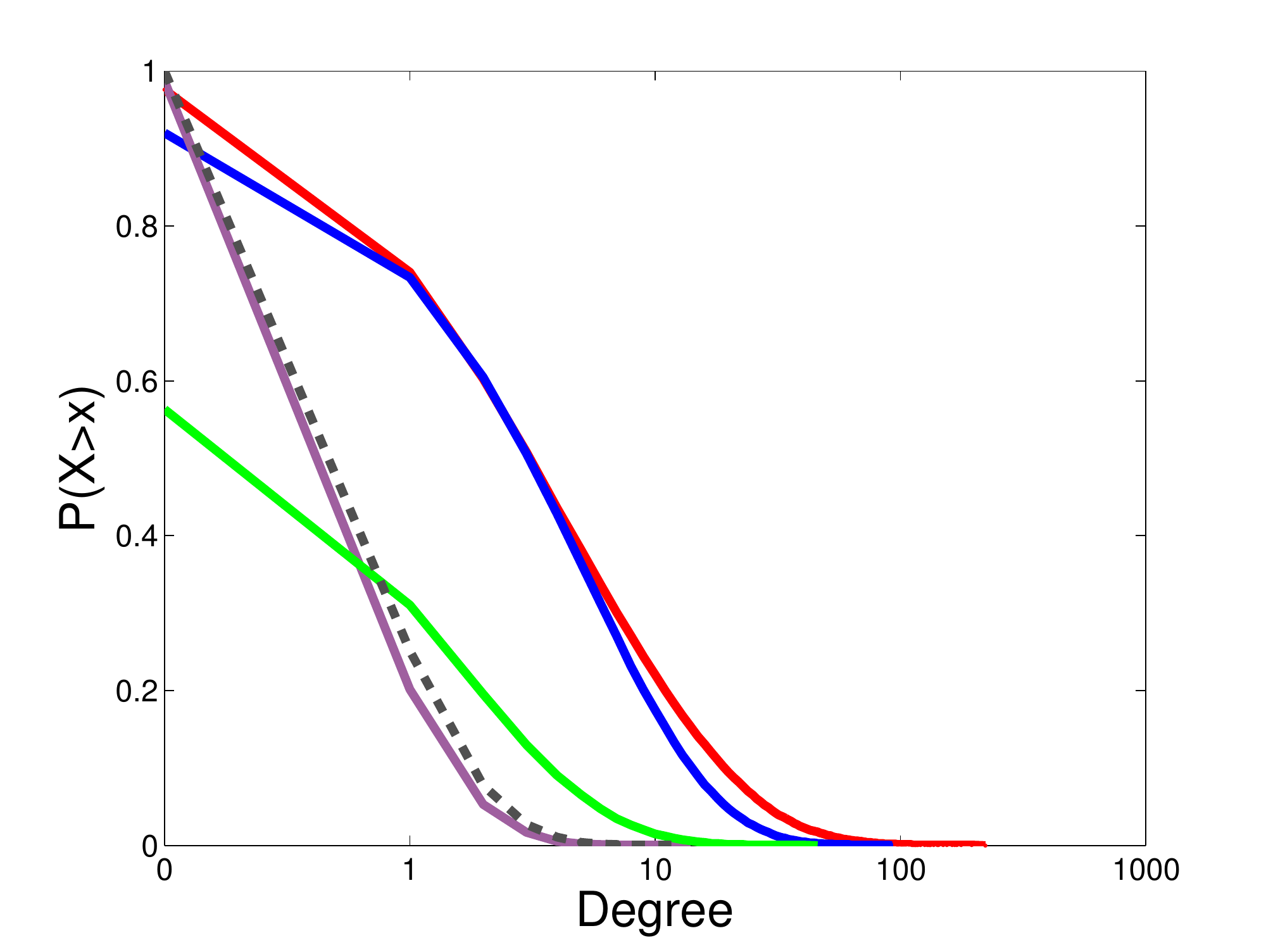}}
\hspace{-5.mm}
\subfigure[Path Length]{\label{fig:fbor pl dist}\includegraphics[width=0.265\textwidth]{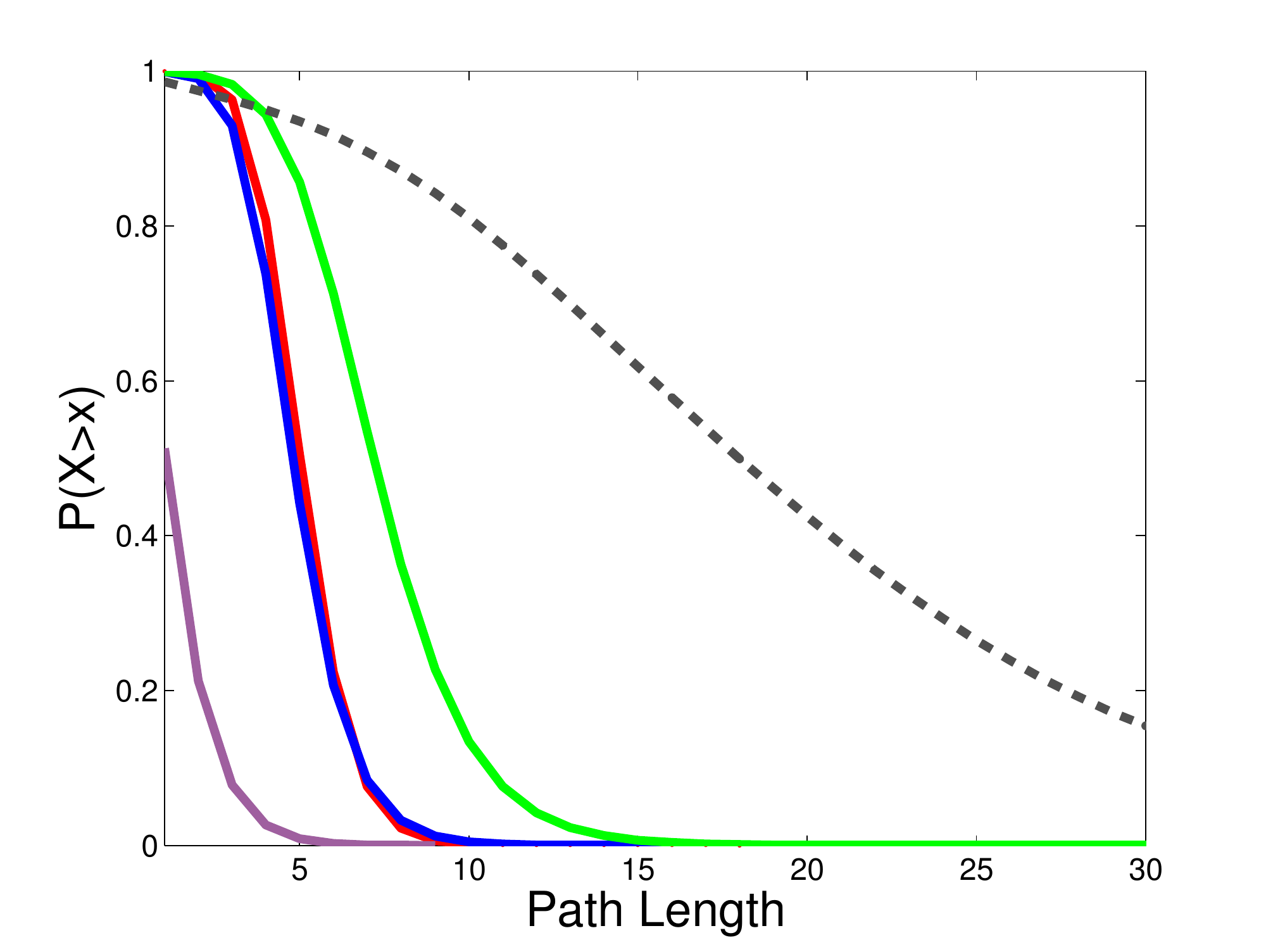}}
\hspace{-5.mm}
\subfigure[Clustering Coef.]{\label{fig:fbor cc dist}\includegraphics[width=0.265\textwidth]{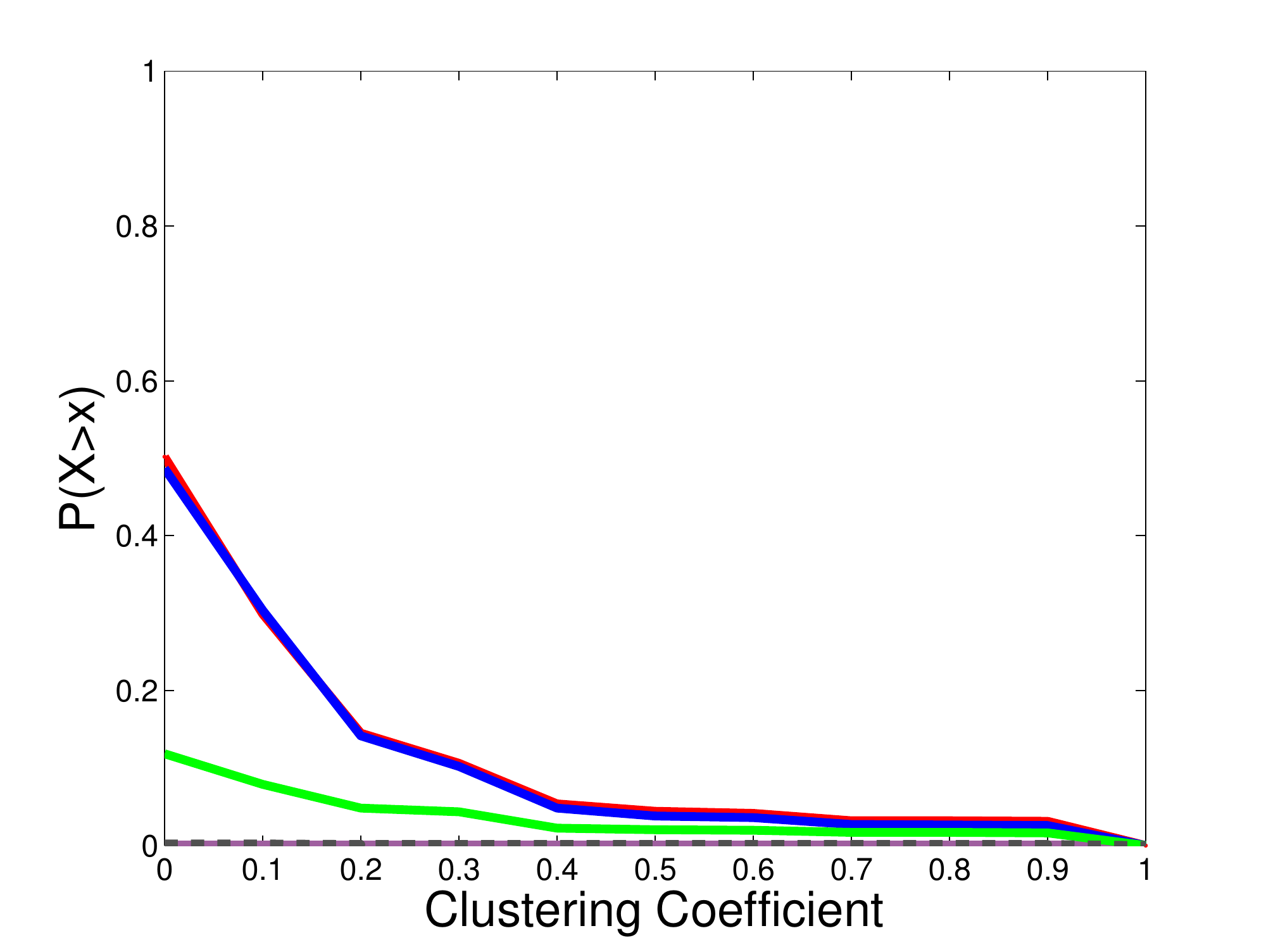}} \\
\subfigure[Degree]{\label{fig:emailpu deg dist}\includegraphics[width=0.265\textwidth]{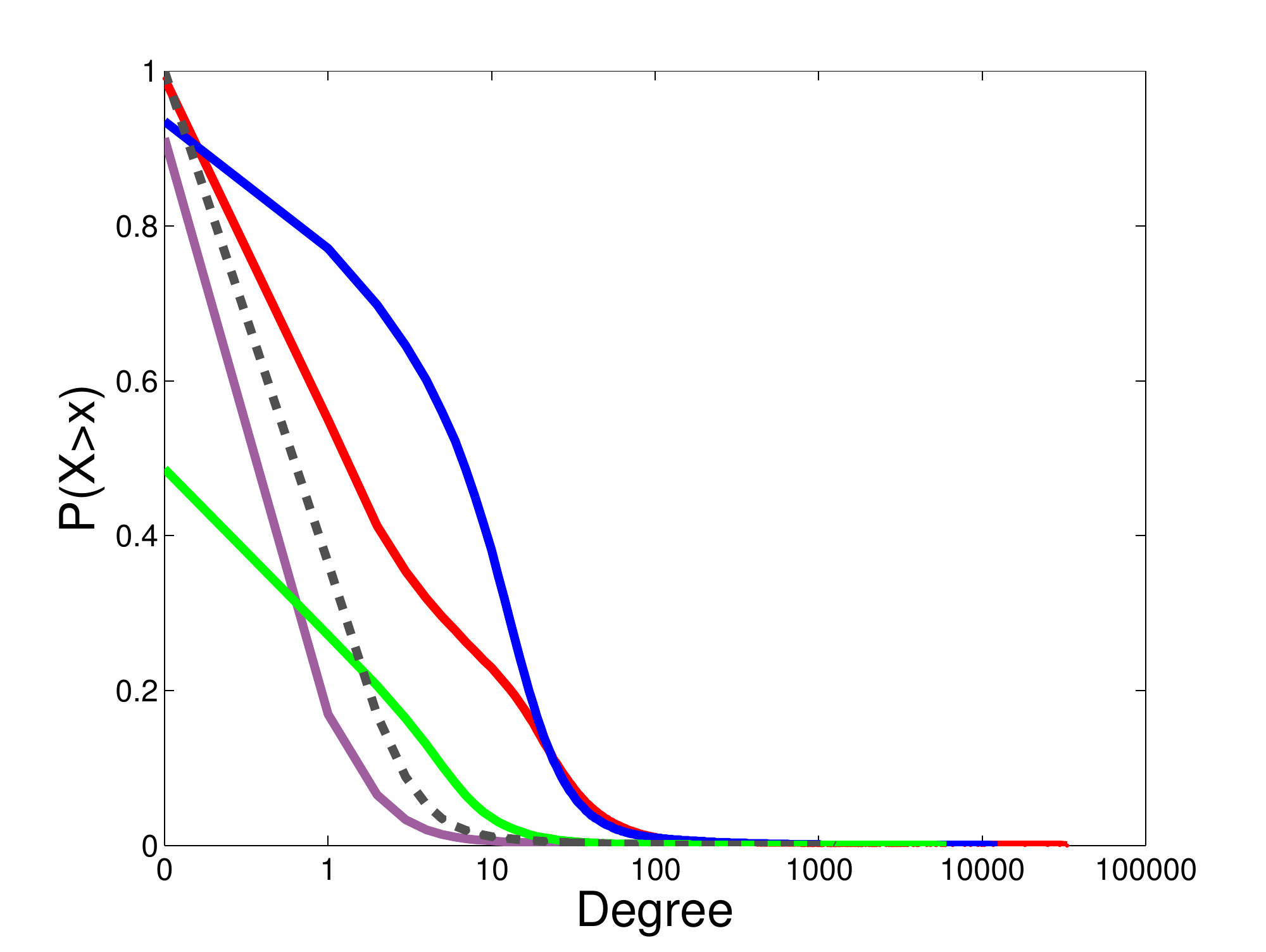}}
\hspace{-5.mm}
\subfigure[Path Length]{\label{fig:emailpu pl dist}\includegraphics[width=0.265\textwidth]{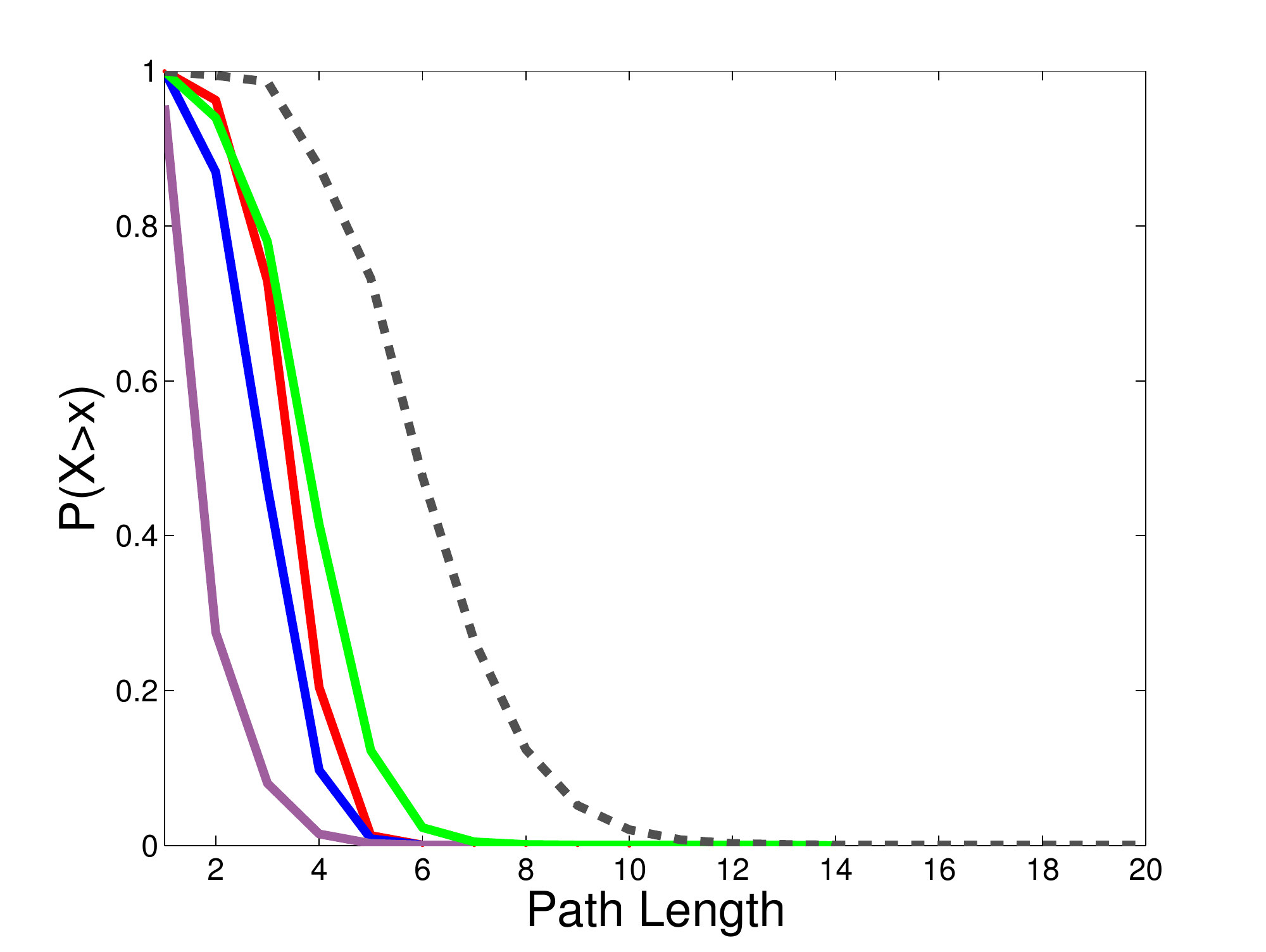}}
\hspace{-5.mm}
\subfigure[Clustering Coef.]{\label{fig:emailpu cc dist}\includegraphics[width=0.265\textwidth]{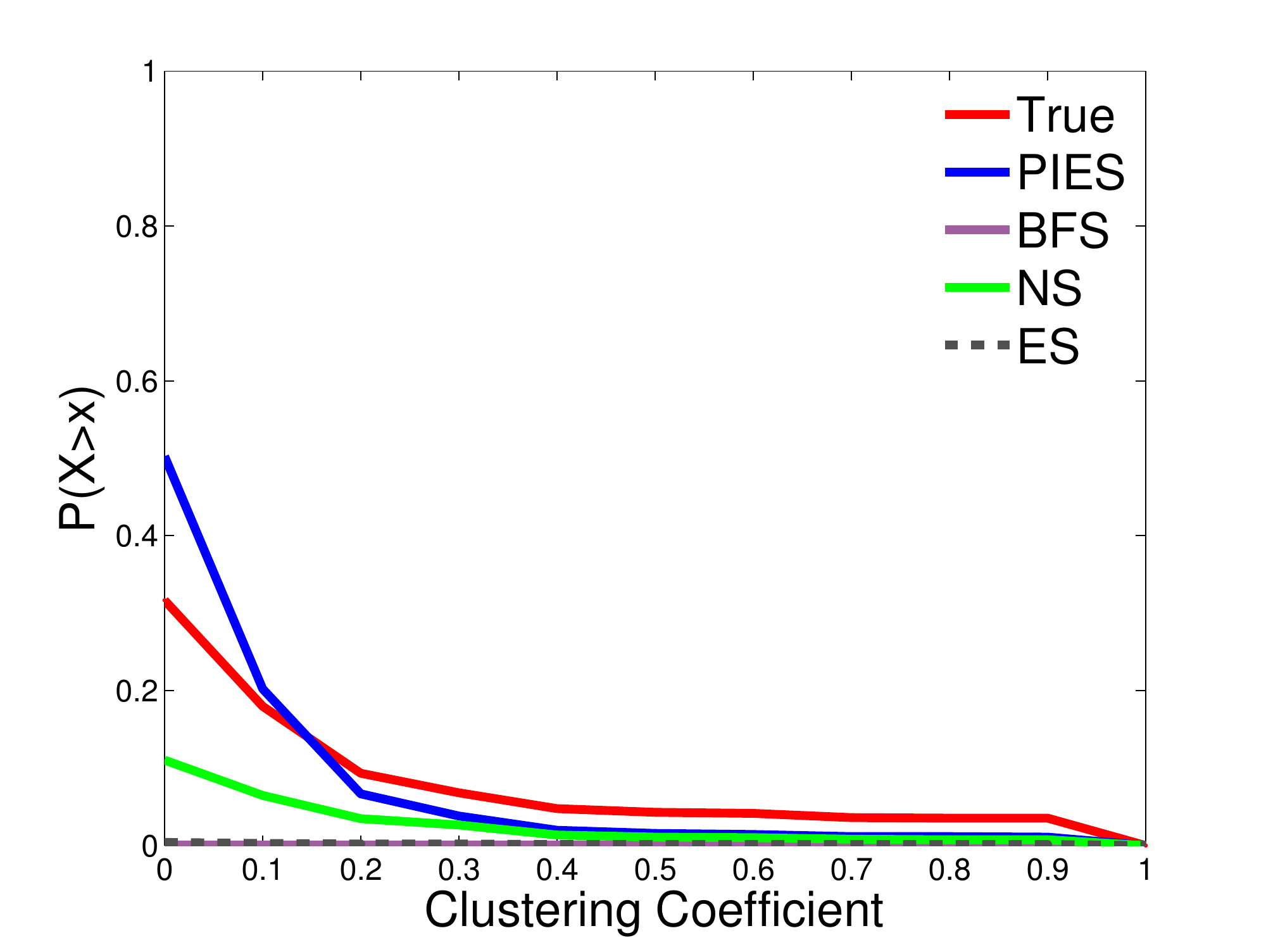}}
\vspace{-2.mm}
\caption{Distributions at 20\% sampling fraction: (a-c)Facebook New Orleans , (d-f)Email Purdue University.}
\label{fig:dist_compare}
\end{figure*}

\paragraph{KS-statistic}
We compute the average of each of these measures
across the five datasets and ten runs for each metric.
Figures~\ref{fig:avg ks deg}--\ref{fig:avg ks comp} show the average KS-statistic  for degree, path length, clustering coefficient and size of connected
components, respectively. We observe that PIES outperforms BFS, NS, and ES for degree, path length, and clustering coefficient. NS comes in the second rank after PIES for the aforementioned measures. Both BFS and ES outperform PIES and NS on the size of connected
components, but they  do not perform well on the other measures. Overall, all sampling algorithms that include
an induced graph step (PIES and NS) in their process perform well for the cases of degree, path length and clustering coefficient as they capture more edges between the sampled nodes.

\paragraph{Distributions} We plot the distributions of the three metrics in
Figure~\ref{fig:dist_compare} for Facebook (a-c), and Email Purdue university (d-f) at 20\% sampling fraction.
We picked the 20\% sampling fraction as a reasonable sample size to show the
difference between the distributions of different sampling algorithms. However, other sampling proportions show similar relative behavior among the algorithms. 

Figures \ref{fig:fbor deg dist} and \ref{fig:emailpu deg dist} show the degree distribution for the two
networks.  From the figures, we can observe that NS under-estimates the degree
of the nodes, resulting in a large fraction of zero-degree (low-degree) nodes in
its sample across the two networks.  Similarly, BFS and ES also capture a large fraction of low-degree nodes.

Figures~\ref{fig:fbor pl dist} and \ref{fig:emailpu pl dist} show the path length distribution for the
two networks. we observe NS samples have a high fraction of long path lengths compared to PIES since it samples low-degree nodes more than high-degree nodes.  

Figures~\ref{fig:fbor cc dist} and \ref{fig:emailpu cc dist} show the clustering coefficient distributions. Across the two networks, NS, ES and BFS produce unclustered samples. PIES performs well for both Email and Facebook networks, however, it performs similar to NS on the HepPH network.

Overall, PIES is the closest to preserving the three distributions compared to other methods. This is due to the fact that PIES samples high degree nodes with a larger probability than NS. Similar to PIES, both BFS and ES select high degree nodes with a probability higher than NS. However, PIES outperforms BFS and ES since it adds extra edges between the sampled nodes (i.e. through partial induction in the forward direction).

We omitted the plots for the size of weakly connected components due to the limited space, however, ES and PIES outperformed the other methods.

\begin{figure}[t]
\centering
\vspace{-2.mm}
\includegraphics[width=0.45\textwidth]{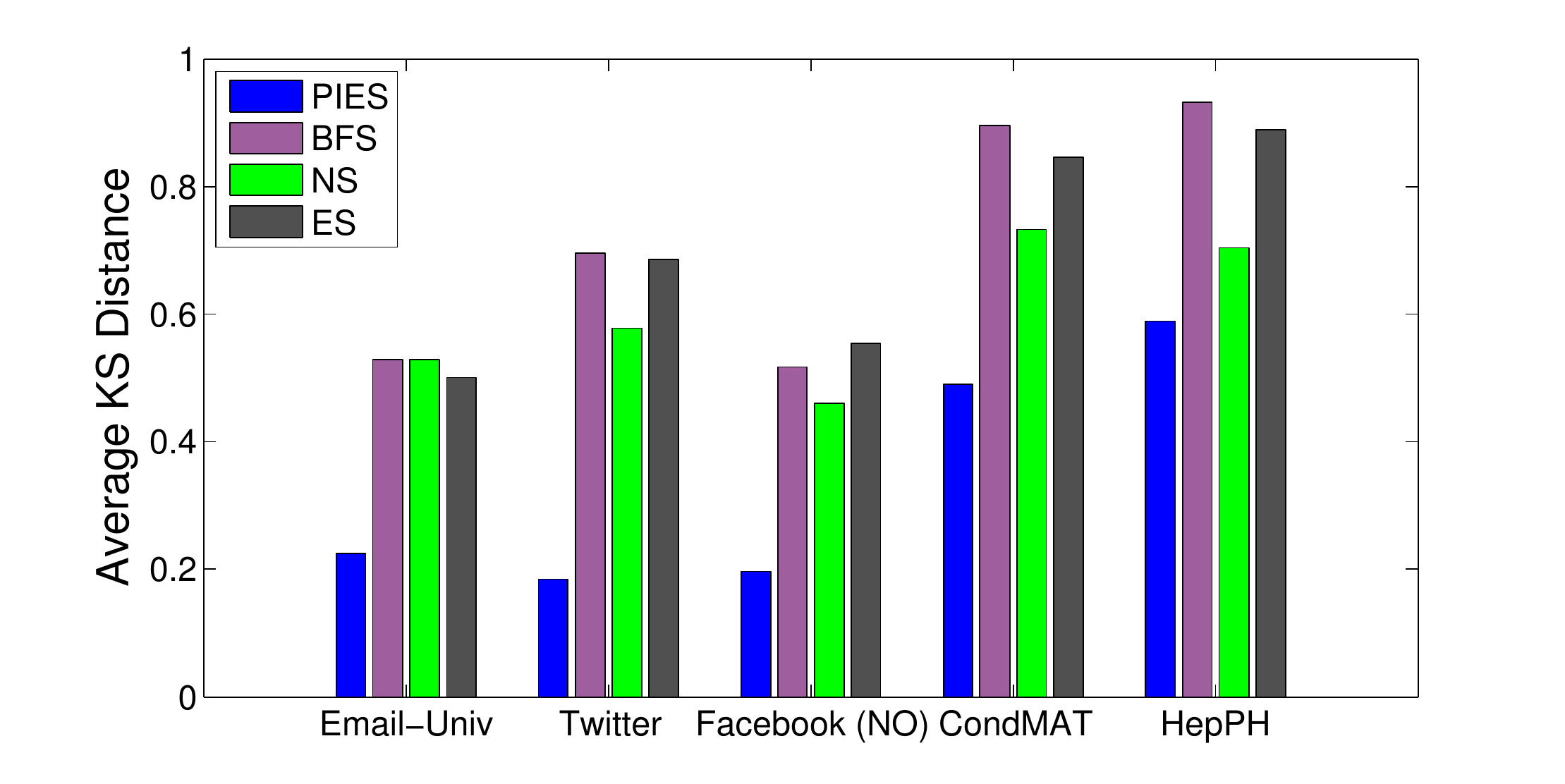}
\vspace{-3.mm}
\caption{Average KS Statistics for different networks (sorted  in increasing order of clustering/density from left to right).}
\label{fig:ks_stats}
\end{figure}

In addition to analyzing the KS statistic as an average on all networks, we also analyze the performance of PIES for each network in Figure~\ref{fig:ks_stats} (average over all graph properties), sorting the networks in increasing order from left to right in terms of their density and clustering. The results indicate that PIES performs better in datasets that are less dense/clustered. This is an interesting result that shows PIES will be more suitable to sample rapidly changing graph streams that are more likely to have a lower density over time.

\begin{figure}[t]
\centering
\vspace{-2.mm}
\hspace{-2.mm}
\subfigure[HepPH]{\label{fig:arxiv stream}\includegraphics[width=0.247\textwidth]{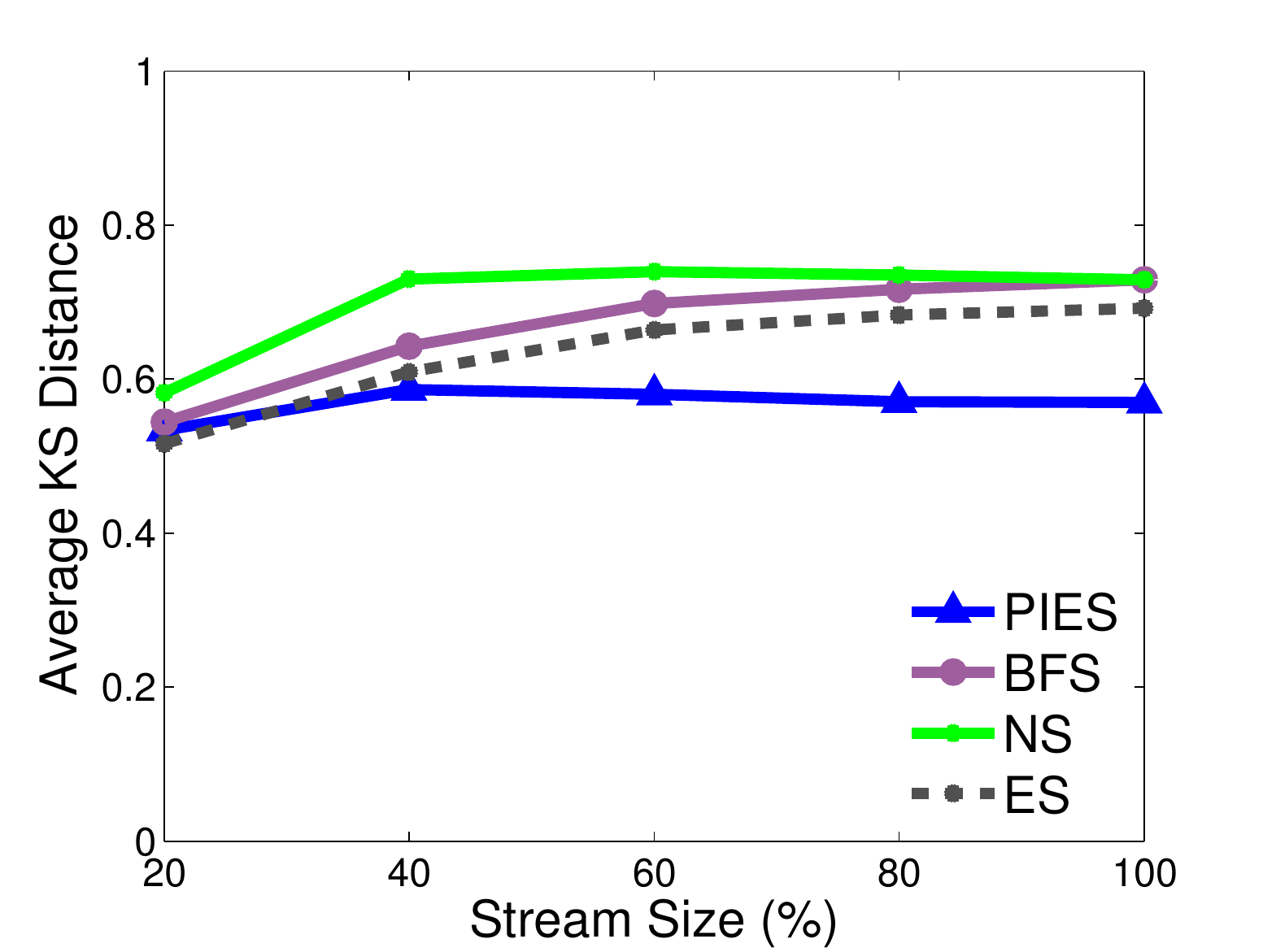}}
\hspace{-4.mm}
\subfigure[Facebook]{\label{fig:fbor stream}\includegraphics[width=0.247\textwidth]{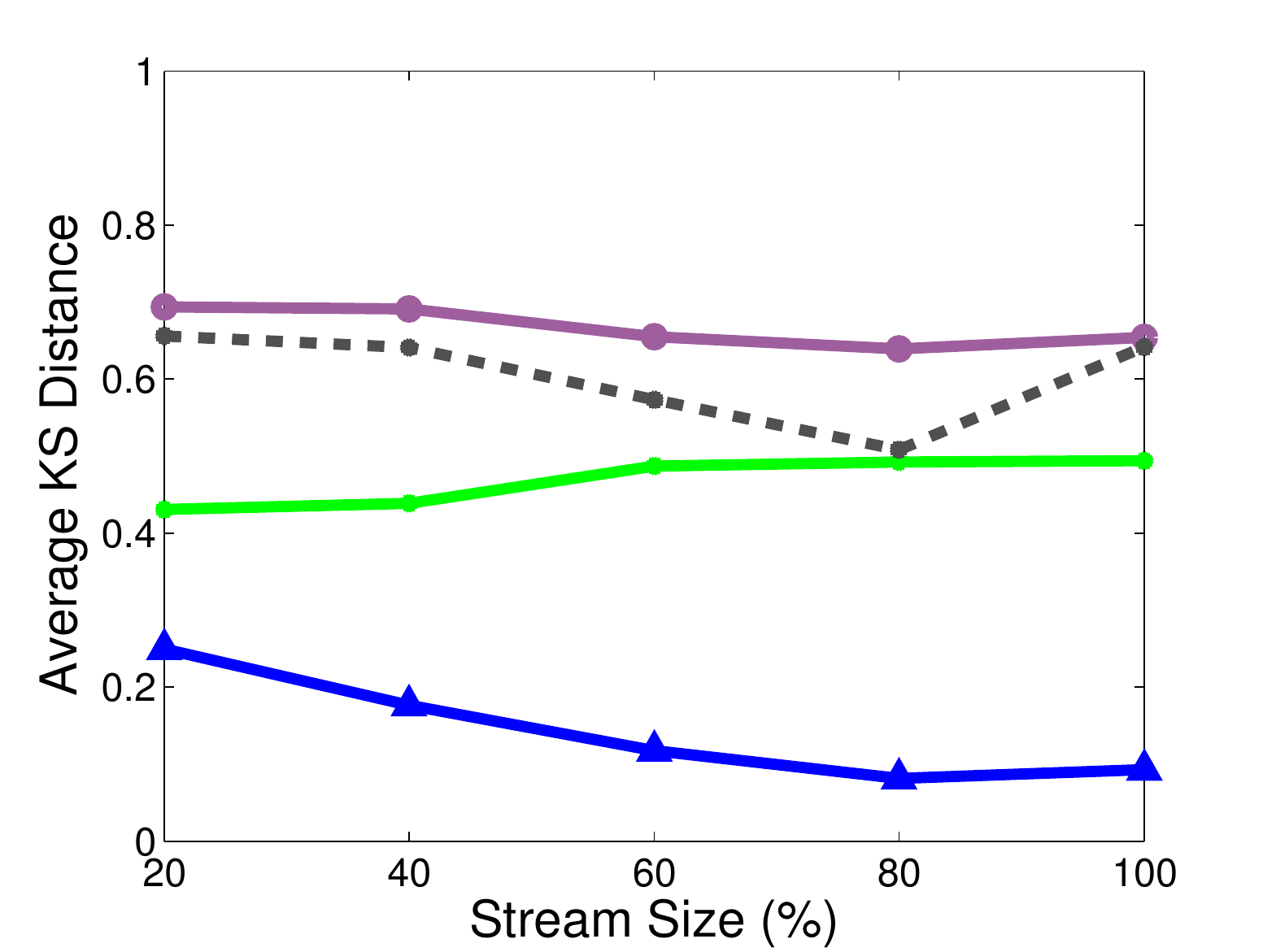}}
\vspace{-3.mm}
\caption{Average KS Statistics at different points of the stream}
\vspace{-3.mm}
\label{fig:dist_compare}
\end{figure}

\paragraph{Evaluation on different points of the stream}
Further, Figures~\ref{fig:arxiv stream}, and \ref{fig:fbor stream} show the KS statistics (average over all graph properties) of the different algorithms at different points in the stream while it is progressing. PIES performs better than NS, BFS, and ES on Facebook. However, PIES performs slightly better than other methods on HepPh. This also illustrates that PIES can maintain a consistently good random sample at different lengths of the stream. 

\begin{figure}[t]
\centering
\vspace{-2.mm}
\includegraphics[width=0.40\textwidth]{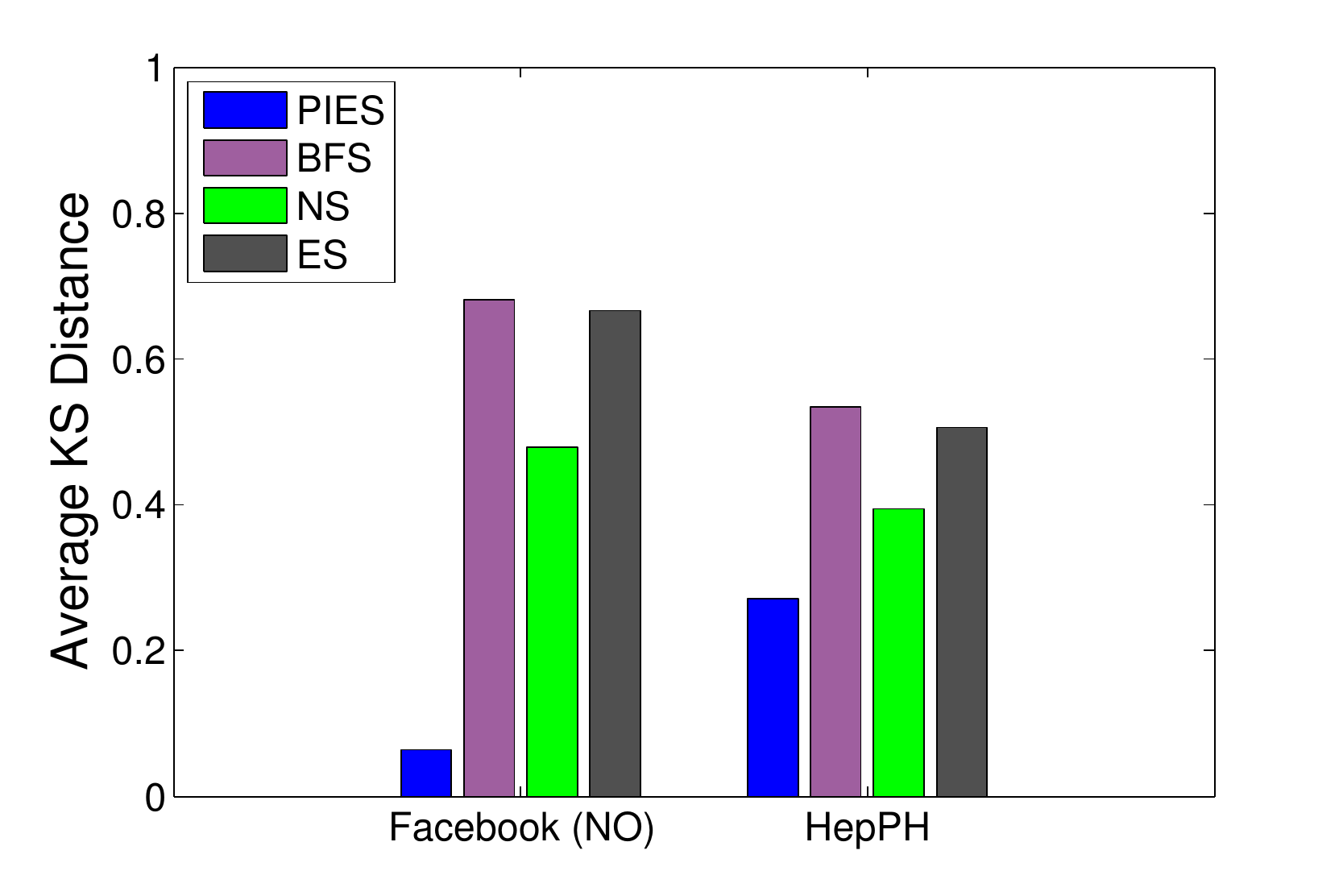}
\vspace{-3.mm}
\caption{Average KS Statistics when the goal is to match the graph back-in time at 20\% of the stream}
\label{fig:ks_back}
\end{figure}

\paragraph{Back-in Time Goal}
Leskovec \etal proposed the back-in time sampling goal \cite{leskovec2006slg} which corresponds to traveling back in time and capturing properties of the past versions of $G$ at sizes $n' < n$. In this experiment, we investigate the question whether we can sample in a manner that allows us to match what the stream looked like in the past. This can help in studying the stationarity properties of the graph stream as it evolves over the time. Figure~\ref{fig:ks_back} shows the average KS statistics (average over all graph properties) of the different algorithms when the goal is to approximate the graph stream back-in time when it was only 20\% the size of the full stream.  We again observe that PIES performs better than the other algorithms. We show the results only for Facebook and HepPh networks, however the same conclusions apply for the other datasets. 

\begin{figure*}[t!]
\centering
\vspace{-2.mm}
\subfigure{\label{fig:flickr deg dist}\includegraphics[width=0.265\textwidth]{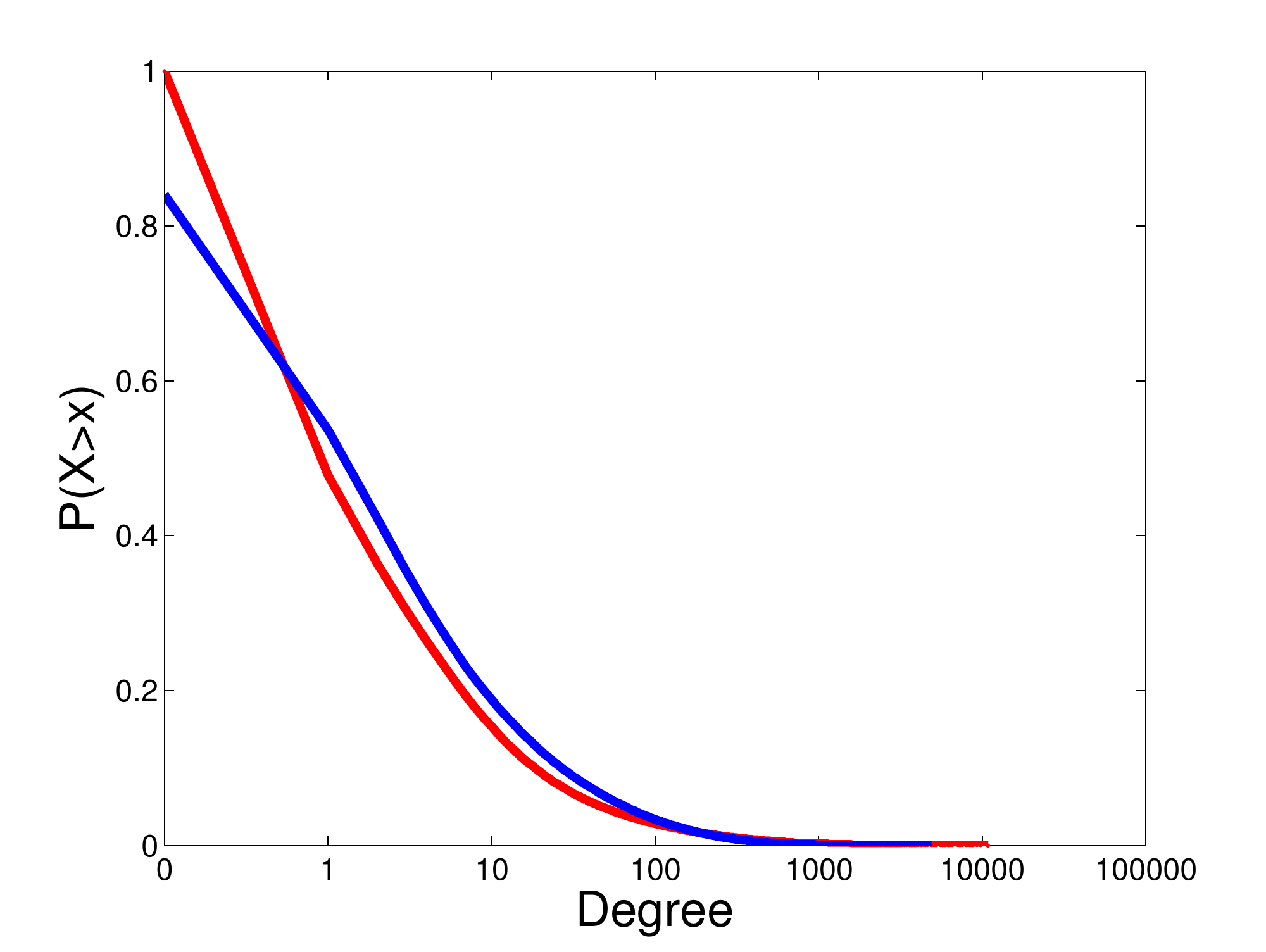}}
\hspace{-5.mm}
\subfigure{\label{fig:flickrpl dist}\includegraphics[width=0.265\textwidth]{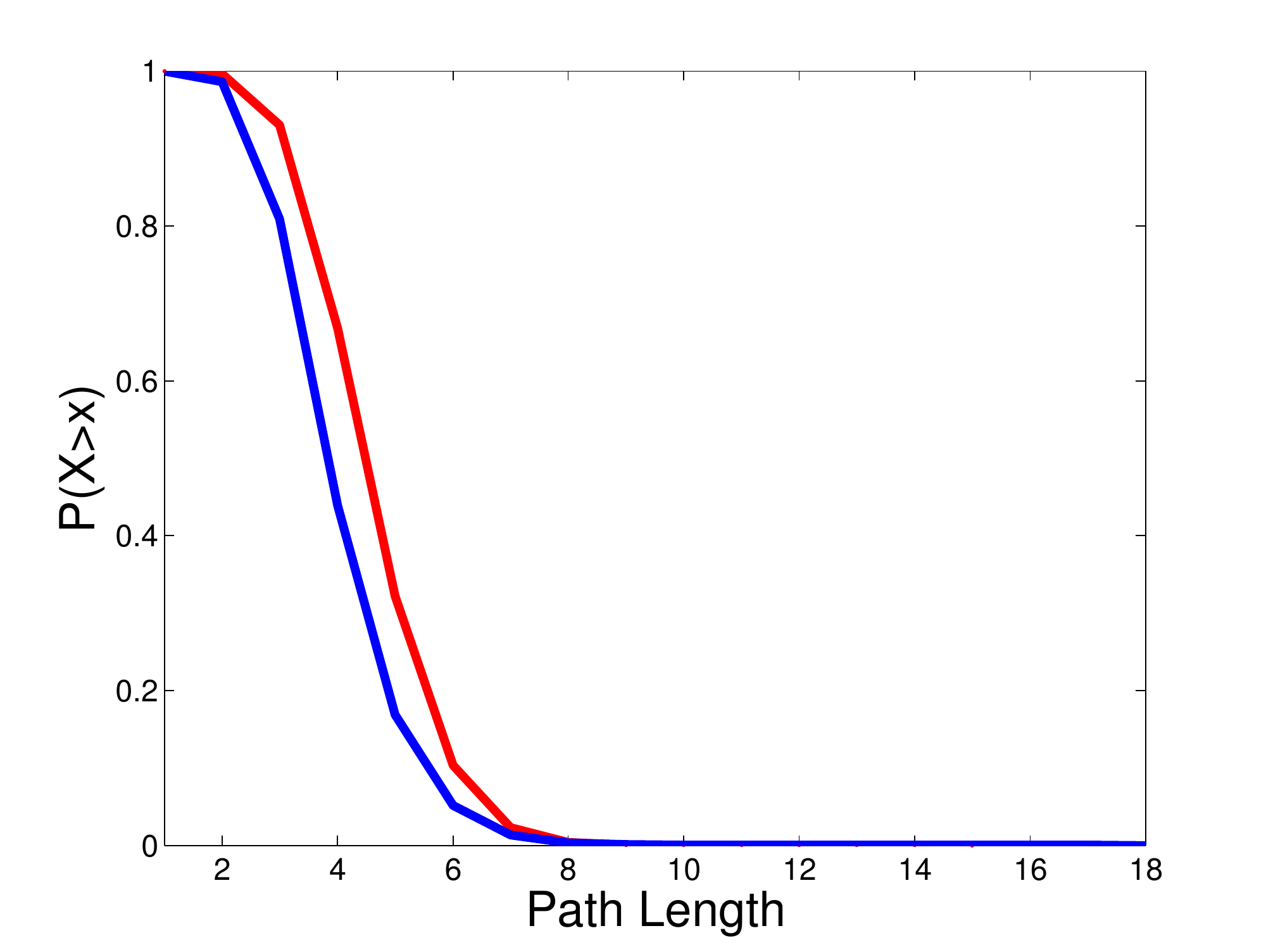}}
\hspace{-5.mm}
\subfigure{\label{fig:flickr cc dist}\includegraphics[width=0.265\textwidth]{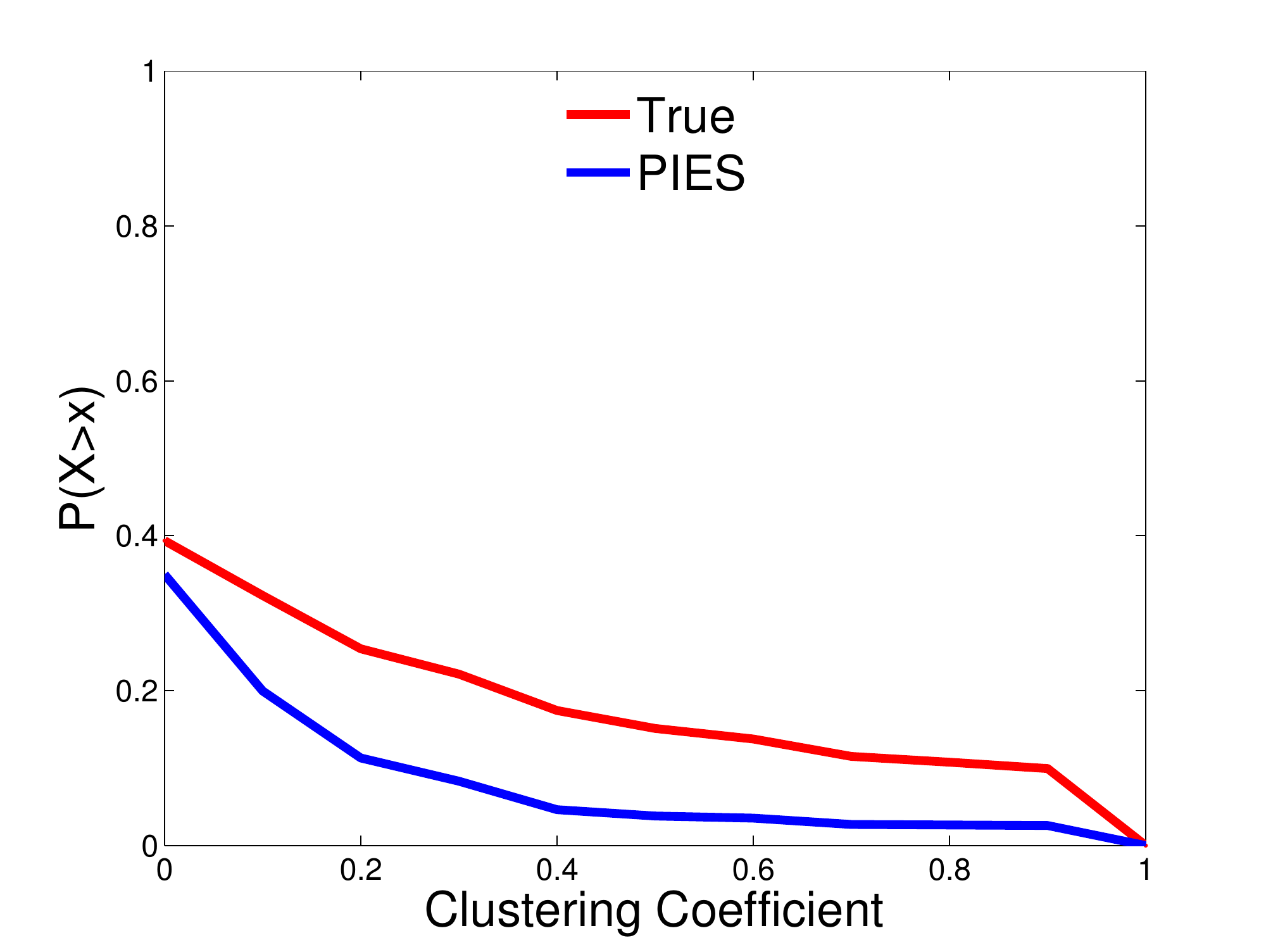}} \\
\vspace{-1.mm}
\caption{Distributions at 30\% sampling fraction for Flickr Network}
\label{fig:dist_flickr}
\end{figure*}

\paragraph{Sampling from Very Large graphs}
While sampling from small graphs (\ie with thousands of nodes/edges) is important for many applications, it is unrealistic for many other applications that deal with very large graphs with hundreds of thousands of nodes/edges. These large graphs are typically too big to fit into memory and therefore they are hard to process with existing sampling methods. Therefore, we also verified our proposed algorithm PIES on large scale graphs with 800,000 nodes and 6,6 million edges collected from Flickr network. As shown in figure~\ref{fig:dist_flickr}, PIES sampled graphs are close to the properties of the larger Flickr network using only {\em a single} pass on the edges \footnote[1]{Note that for the Flickr data experiments, we compare PIES to the true distribution only, since the other baseline methods are inefficient to run for very large graphs and they don't match the graph properties well on smaller sampling sizes.}.

\paragraph{Comparison with non-streaming algorithms}
Our goal is to obtain a representative sample from a stream that is either evolving over the time or too large to fit into memory. Here we compare PIES to other non-streaming sampling algorithms. We compare to Forest Fire Sampling (FFS)~\cite{leskovec2006slg} and fully-induced edge sampling (ES-i). In the case of FFS, we use $p_f=0.7$ as in \cite{leskovec2006slg}. In ES-i, we first sample the edges with ES then we add all the edges among the sampled nodes in a second pass (full induction). 

\begin{figure}[t]
\centering
\vspace{-2.mm}
\includegraphics[width=0.45\textwidth]{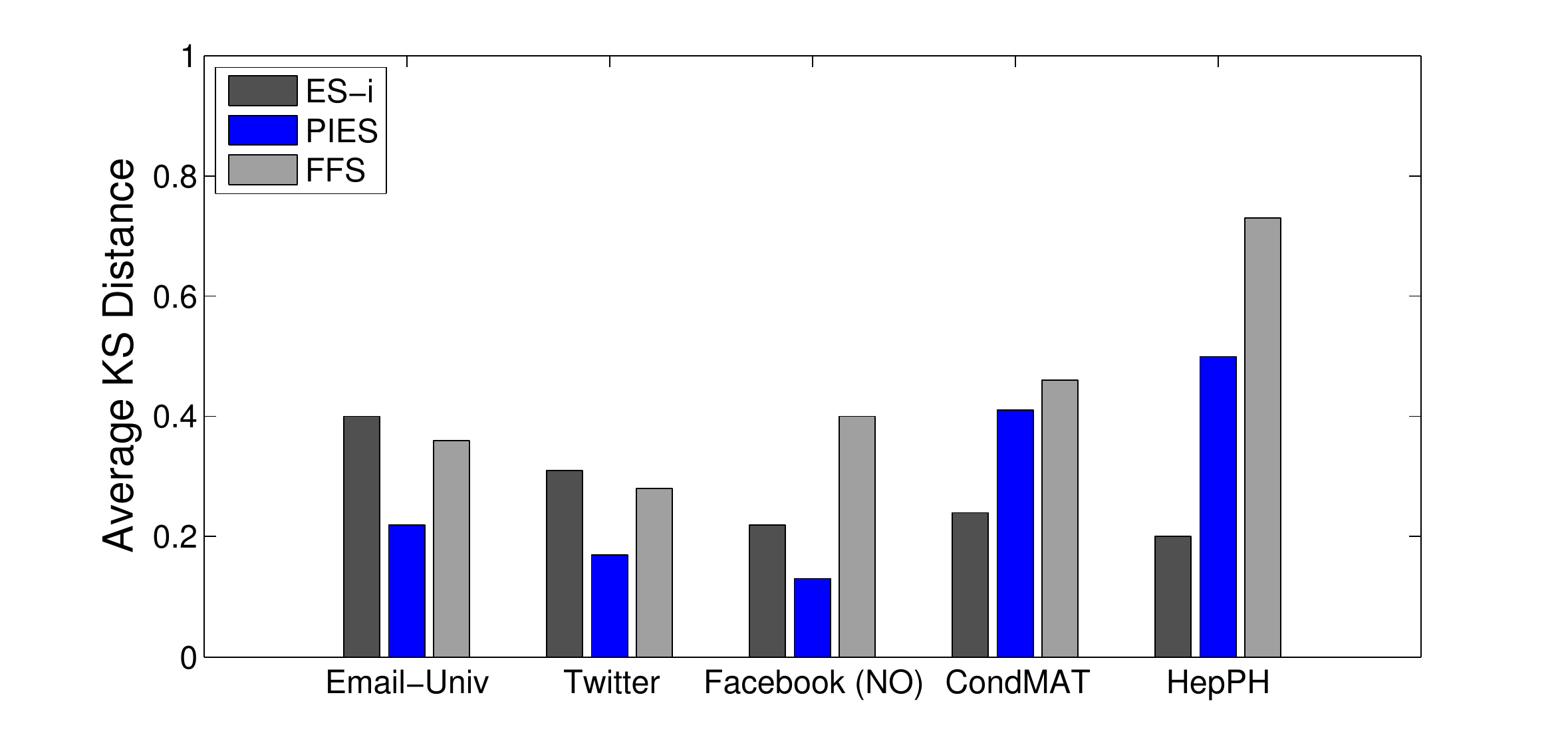}
\caption{Average KS Statistics for different networks (sorted in increasing order of clustering/density from left to right).}
\label{fig:ks_nonstream}
\end{figure}

Figure~\ref{fig:ks_nonstream} shows the average KS statistic (average over all graph properties) for the five networks. Overall, PIES performs better than both ES-i and FFS. However, ES-i performs better for HepPh and ConMAT. This illustrates the effect of full induction versus partial induction for more dense networks. Since ES-i gets the chance to add more edges among the sampled nodes, it outperforms PIES on graphs with higher density/clustering. However, PIES performs better for the less dense, and clustered graphs.  

\section{Related work}
\label{sec:related}
\vspace{-1.mm}
\paragraph{Sampling from Graphs} The problem of sampling graphs has been of interest in many different fields of
research. The work in \cite{lee:06,yoon08,stumpf2005ssf} studies the statistical
properties of samples from complex networks produced by traditional sampling
algorithms such as node sampling, edge sampling and random-walk based sampling and
discusses the biases in estimates of graph metrics due to sampling.  The work in \cite{Maiya2011kdd} also discusses the 
connections between specific biases and various measures of structural representativeness.
In addition, there have been a number of sampling algorithms in other communities such as 
in peer-to-peer networks \cite{stutzbach2006usu,gkantsidis2004rwp}.
Internet modeling research community~\cite{krishnamurthy2007sli}
and the WWW information retrieval community has focussed on random walk based
sampling algorithms like PageRank ~\cite{page1998pagerank,kleinberg1999authoritative}. 
There is also some work that highlights the different aspects of the sampling problem. Examples include \cite{kolaczyk2009statistical,avrachenkov2010improving,ahmed2010reconsidering}

In social networks research, the recent work in \cite{ribeiro10imc} uses random walks to
estimate node properties in $G$ (e.g., degree distributions in online social
networks).  These different sampling algorithms focused on estimating
either the local or global properties of the original graph, but {\em not} to
sample a representative subgraph of the original graph, which is our goal. The work in \cite{Maiya2010www} studied the problem of 
sampling a subgraph representative of the graph community structure by sampling the nodes that maximize the expansion. 

Due to the popularity of online social networks such as Facebook
and Twitter, there has been a lot of
work~\cite{mislove07imc,leskovec08wbp,leskovec08kdd,kumar06kdd,ahn2007analysis,chun2008comparison}
studying the growth and evolution of these networks. While most of
them have been on static graphs, recent
works~\cite{wilson09eurosys,viswanath-2009-activity} have started focusing on
interactions in social networks. There is also work on decentralized search and crawling \cite{baykan2009comparison,gjoka2010walking,kurant2011towards}, however, in our work we focus on sampling from graphs that are naturally evolving as a stream of edges.
In the literature, the most closely related efforts are that of Leskovec \etal in
\cite{leskovec2006slg} and Hubler \etal in \cite{hubler08icdm}. But, as we mentioned before, our work is different as we focus on the novel problem of sampling from graphs that are naturally evolving as a stream of edges (graph streams).

\paragraph{Impact of Sampling on Other applications} Recently, some research has also focused on how the different sampling methods impact the performance of applications overlaid on the networks. One such study investigated the impact of sampling designs on the discovery of the information diffusion process \cite{de2010does}. Another study investigated the impact of the choice of the sampling design on the performance of relational classification algorithms \cite{ahmed2012network}.

\paragraph{Data and Graph Streams} Although significant work has been proposed to solve the problem of graph sampling, to our knowledge, there is no prior research on sampling from graph streams to obtain a representative subgraph. 
However, several research works \cite{atish,yossef,guha} studied graph streaming algorithms for counting triangles, degree sequences, and estimating page ranks. The main contributions of these works are to use a small amount of memory (sublinear space) and few passes to perform computations on large graphs streams. In database research, some research studied data stream management systems. For example, the work in \cite{manku} studied the problem of computing frequency counts in data streams, and the work in \cite{agrwal} studied the problem of sampling from data stream of database queries.  

\section{Conclusions}
\label{sec:conclusions}
\vspace{1.mm}
Much of the past efforts on sampling networks have assumed
that the sampling algorithm can access the full graph in
order to decide which nodes/edges to select.  However, many
large-scale network datasets are constructed from a graph
{\em stream} consisting of micro-communications among users
(\eg wall posts, tweets, emails). In this work, we have
formulated the problem of sampling {\em representative}
subgraphs from such large graph streams.  We proposed a
novel sampling algorithm, PIES, that is based on combining
edge sampling with partial induction. Our approach is not
only simple and efficient, it is also amenable to a
streaming implementation. Furthermore, our empirical results
show that PIES significantly outperforms other sampling
algorithms, both streaming and non-streaming, across a range
of real-world network datasets. In future work, we aim to study the theoretical properties of graph stream sampling in particular our proposed method.
\vspace{-3.mm}

\section{Acknowledgements}
\begin{small}
This research is supported by ARO, NSF under contract
number(s) W911NF-08-1-0238, IIS-1017898, IIS-0916686, IIS-1149789.
The views and conclusions contained herein are
those of the authors and should not be interpreted as necessarily
representing the official policies or endorsements either expressed
or implied, of ARO, NSF, or the U.S. Government.
\end{small}
{
\vspace{1.mm}
\scriptsize
\bibliographystyle{abbrv}
\bibliography{paper}
}

\end{document}